\documentclass[11pt]{article}

\usepackage[preprint]{acl}
\usepackage{booktabs}
\usepackage{multirow}
\usepackage{multicol}
\usepackage{times}
\usepackage{latexsym}

\usepackage[T1]{fontenc}

\usepackage[utf8]{inputenc}

\usepackage{microtype}

\usepackage{inconsolata}

\usepackage{graphicx}

\title{Label Over Logic? \\How Source Cues Bias Human Fallacy Judgments More Than LLMs}

\author{
 \textbf{Mahjabin Nahar},
 \textbf{Nafis Irtiza Tripto},
 \textbf{Aiping Xiong},
 \\
 \textbf{Ting-Hao \lq Kenneth\rq \space Huang},
 \textbf{Dongwon Lee}
\\
 The Pennsylvania State University, USA
 \\
 \texttt{\{mahjabin.n, nit5154, axx29, txh710, dongwon\}@psu.edu} \\
}

\begin{document}
\maketitle
\begin{abstract}
As AI-generated and AI-assisted content floods online spaces, source labels attached to such content can distort human reasoning judgments, with downstream consequences for moderation, evaluation, and decision-making. Whether LLMs share this vulnerability, or offer more source-agnostic evaluation, remains an open question with strong implications for human-AI collaboration. We examine this issue using \textit{logical fallacies} as a controlled setting to isolate source-label effects on reasoning quality, independent of domain knowledge. We conduct an online study ($N=505$) where participants are assigned to a source condition (human, AI, human with AI assistance, AI with human assistance, or no disclosure) and evaluate comments containing logical fallacies, comparing their judgments with those of LLMs (GPT-5.2, Gemini 2.5 Flash, Claude Sonnet 4.5), which were evaluated across the same source conditions. Human evaluators were significantly more susceptible to fallacies labeled as \lq written by human\rq \space or \lq written by human with AI assistance\rq \space and assigned higher trust ratings in these conditions. LLM evaluations remained comparatively stable across source labels, though performance varied across models. Confidence levels were similarly high across conditions for both humans and LLMs, regardless of the presence of fallacies. Our findings indicate that source-label bias is primarily a human vulnerability for logical fallacy evaluation, with potential implications in human-LLM collaboration in increasingly AI-mediated environments\footnote{All code, data, and study materials are available at: \href{https://github.com/MahjabinNahar/Label_Over_Logic}{https://github.com/MahjabinNahar/Label\_Over\_Logic}}. 
\end{abstract}

\section{Introduction}
\begin{figure}[t]
\centering
  \includegraphics[width=\columnwidth]{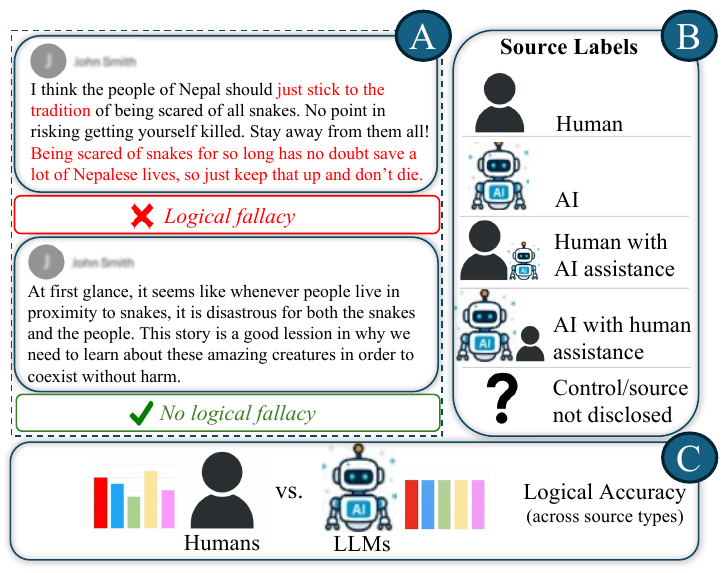}
  \caption{(A) We select news comments with and without logical fallacies from the CoCoLoFa dataset \cite{yeh2024cocolofa}. The illustrated example shows an \textbf{appeal to tradition fallacy, where a claim is justified using tradition as evidence}, e.g., the comment argues that \textcolor{red}{being scared of snakes for so long has saved a lot of Nepalese lives}, so people should \textcolor{red}{just stick to the tradition} to stay alive. (B) We evaluate the impact of source perception by presenting these comments with one of five source labels in a between-subjects design. (C) We compare human and LLM evaluations to assess susceptibility to source bias in logical reasoning.
  }
  \label{fig:teaser}
\end{figure}

Since ChatGPT was introduced in 2022, online spaces have been flooded with AI-generated and AI-assisted content~\cite{sun2025we}. 
While platforms such as Instagram have experimented with AI-generated comments to enhance user engagement~\cite{marketingtechnews}, Reddit has taken the opposite stance by banning researchers who deployed AI bots to influence discussions~\cite{theverge}, highlighting a broader tension in how AI-mediated content is simultaneously embraced for its usefulness and scrutinized for its potential to manipulate users.

It is important to consider how such content is presented and labeled, as \textit{source labels} can influence human judgment in mixed and sometimes counterintuitive ways that may not align with content quality~\cite{rae2024effects,sun2026label,zhang2023trust}. AI-generated content can also be highly persuasive, often outperforming humans in producing misleading yet convincing arguments~\cite{goldstein2024persuasive,salvi2025conversational}. When such content is paired with misleading or strategically framed source labels~\cite{zhu2025human}, users may place trust in content that does not align with its underlying quality or origin. Malicious actors may exploit such labels to sway judgment, while unsuspecting users further propagate misinformation under misleading source cues. These biases can distort downstream processes such as content moderation and evaluation, consistent with broader evidence of context-dependent bias in human decision-making~\cite{dror2020cognitive}.

As large language models (LLMs) increasingly demonstrate strong performance on reasoning tasks~\cite{zheng2023judging}, it is critical to examine whether they exhibit \textit{source-related biases}. If not, they may serve as complementary tools in decision-making~\cite{kumar2024watch, liu2023g, park2025llm}, potentially mitigating human biases. On the other hand, LLMs are also known to exhibit biases \cite{sclar2024quantifying, sun2026label} and may mirror those observed in humans.

At the same time, online discourse is frequently shaped by logical fallacies. Logical fallacies are deceptive reasoning errors that undermine argument validity \cite{stanfordencyclopedia, walton1987informal}, degrade discussion quality, and contribute to the spread of misinformation across a wide range of contexts \cite{sahai2021breaking, yeh2024cocolofa}. Detecting logical fallacies has emerged as an important reasoning task as it does not require external factual knowledge and instead relies on evaluating the internal structure of arguments~\cite{jin2022logical}, providing a controlled setting for studying reasoning judgments.

Nevertheless, it remains unclear whether humans and LLMs are similarly susceptible to source-based reasoning biases. As AI assistance takes many forms \cite{zhang2023human}, ranging from human-led to AI-led collaboration, it is important to examine how different source conditions (AI, human, AI with human assistance, human with AI assistance, and no disclosure) influence the evaluation of logical fallacies. If source bias is present, it may affect not only perceived logical accuracy but also participants’ confidence, trust, and  evaluation of the commenter \cite{chen2023ai, nahar2025catch}, raising concerns about judgment reliability in AI-mediated settings. As such, we ask two pivotal questions:

\textbf{RQ1:} How do participants’ and LLMs’ (a) perceived logical accuracy of and (b) confidence in evaluating content with and without logical fallacies vary across perceived source conditions?

\textbf{RQ2:} How do perceived sources influence participants' (a) trust in the commenter and (b) overall evaluation of the commenter?

We conduct a human-subjects study ($N = 505$) on Prolific\footnote{https://www.prolific.com/} using news comments with and without logical fallacies from the CoCoLoFa benchmark~\cite{yeh2024cocolofa}; Figure \ref{fig:teaser}. The comments cover eight common fallacy types in online discourse: \textit{appeal to authority, appeal to majority, appeal to nature, appeal to tradition, appeal to worse problems, false dilemma, hasty generalization, and slippery slope}; Appendix \ref{appendix: fallacy types}. Participants are assigned to a source condition (\textbf{human, AI, human with AI assistance, AI with human assistance, and no disclosure}) and evaluate identical content. We then compare human and LLM judgments (GPT-5.2, Gemini 2.5 Flash, Claude Sonnet 4.5), which were evaluated across the same source conditions. We also examine human and LLM evaluations across fallacy types and how humans interpret hybrid human-AI authorship labels.

\textbf{Humans were more susceptible to fallacies when they were labeled as authored by humans or humans with AI assistance, and they also assigned higher trust ratings in these conditions.} Overall evaluations most clearly favored human-written content, while the advantage of content written by humans with AI assistance was only suggestive among participants who correctly identified the source label. An exploratory analysis showed that human evaluators interpreted \lq\lq human with AI assistance\rq\rq \space and \lq\lq AI with human assistance\rq\rq \space differently: the former as human ideas polished by AI and the latter as AI ideas refined by humans. 

In contrast, \textbf{LLMs remained stable across source labels, though performance varied by model}. Confidence levels were mostly stable across conditions for humans and LLMs, regardless of fallacy presence. Moreover, fallacy-level analyses suggested that humans and LLMs may have complementary weaknesses. Broadly, our findings highlight the value of human–LLM collaboration for decision-making in AI-mediated environments.

\begin{figure*}[t]
\centering
\includegraphics[width=\textwidth]{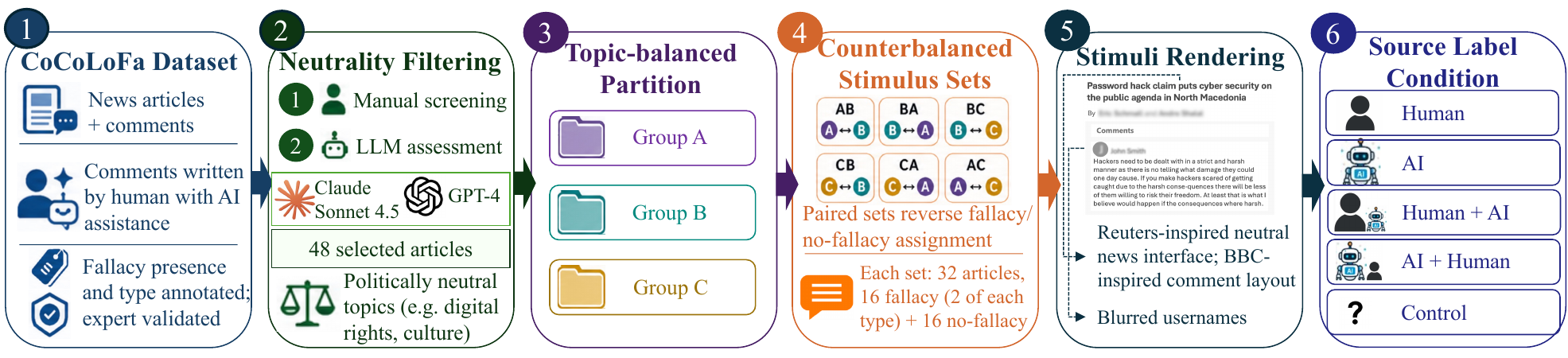}
\caption{An overview of the stimuli generation pipeline.}
\label{fig:stimuli_generation}
\end{figure*}

\section{Related Work}
\textbf{Source perception in human-AI interaction.} Prior work shows that perceived source shapes user judgments, though the nature and strength of these effects vary considerably across tasks and contexts \cite{ gallegos2025labeling, jain2024revealing, li2025nudging, lin2026visible, sun2026label, zhang2023human, zhu2025human}. Source cues also shape downstream perceptions such as trust, credibility, and evaluation scores, often in ways that do not align with content quality \cite{chen2023ai, zhou2025effect}. In content creation, writing tasks, and news settings, AI involvement is frequently penalized \cite{cheong2025penalizing, dai2024ai, jia2024news, li2024does, lim2024effect, rae2024effects}. However, in decision-making and collaborative settings, users may rely on or even prefer AI systems, particularly when they are perceived as more capable or reliable \cite{klingbeil2024trust, logg2019algorithm, zhang2023trust}. These findings suggest that source-label effects are mixed and context-dependent. Prior source-label studies have often examined general content quality, credibility, persuasion, or writing evaluation, whereas their impact on logical fallacies is less understood. Moreover, as AI-assisted authorship becomes increasingly common \cite{stanfordaiindex}, hybrid forms of authorship are becoming more consequential, yet their effects remain underexplored. We address these gaps by examining how human, AI, and hybrid authorship labels shape judgments of otherwise identical arguments with and without logical fallacies.

\textbf{Logical fallacies and reasoning evaluation in NLP.} Logical fallacies have emerged as a useful testbed for studying reasoning quality, attracting growing attention in NLP research \cite{jin2022logical}. Computationally, researchers have built datasets and benchmarks such as CoCoLoFa \cite{yeh2024cocolofa} and MAFALDA \cite{helwe2024mafalda}, and developed methods for fallacy detection \cite{sahai2021breaking, sourati2023robust}. While LLMs perform better with structured prompting and counterargument generation \cite{jeong2025large}, they remain susceptible to certain fallacy patterns \cite{payandeh2024susceptible}. On the human side, research consistently shows that people are imperfect fallacy detectors \cite{hruschka2023learning} and tend to rely on heuristic reasoning rather than formal logic \cite{chen2026emotionally, dror2020cognitive}. Despite this rich body of work, prior research has rarely examined how source perception influences reasoning evaluation across both humans and LLMs, motivating us to study these processes jointly.

\textbf{Biases in LLM reasoning evaluation.} 
While LLMs increasingly demonstrate strong performance on reasoning tasks~\cite{zheng2023judging}, they can also be 
influenced by contextual information beyond the reasoning problem itself. Irrelevant or misleading context can reduce reasoning accuracy \citep{shi2023large, zhao2026thinking}, whereas prompt-level biasing features can steer predictions \citep{turpin2023language}. Prior work further shows that both human and LLM evaluators can be susceptible to perturbations such as misinformation and authority cues \citep{chen-etal-2024-humans}. Therefore, it is important to understand how LLMs evaluate reasoning quality with varying source labels when the underlying content remains constant and how these effects compare with those observed in humans. However, prior work has not examined how such authorship cues affect judgments of logical reasoning in the presence of fallacies, particularly when authorship includes hybrid human-AI configurations.

\section{Methods}


\subsection{Stimuli and Presentation} \label{subsec: stimuli and presentation}
We construct our stimuli using the CoCoLoFa dataset \cite{yeh2024cocolofa}, which contains news comments written by humans with AI assistance, annotated for logical fallacy presence and type. The annotations were validated by expert annotators, making the dataset well-suited for studying reasoning quality. To reduce potential confounds from politically biased or emotionally charged content, we curate 48 news articles and associated comments from relatively neutral domains (e.g., digital rights, culture); Appendix \ref{selected topics and filtering strategy}. We further verify the neutrality of the news titles and comments through a two-step process: manual screening and LLM-based assessment \cite{nahar2024fakes} using Claude Sonnet 4.5 and GPT-4; Figure \ref{fig:stimuli_generation}.



\begin{figure}[hbt!]
\centering
  \includegraphics[width=\columnwidth]{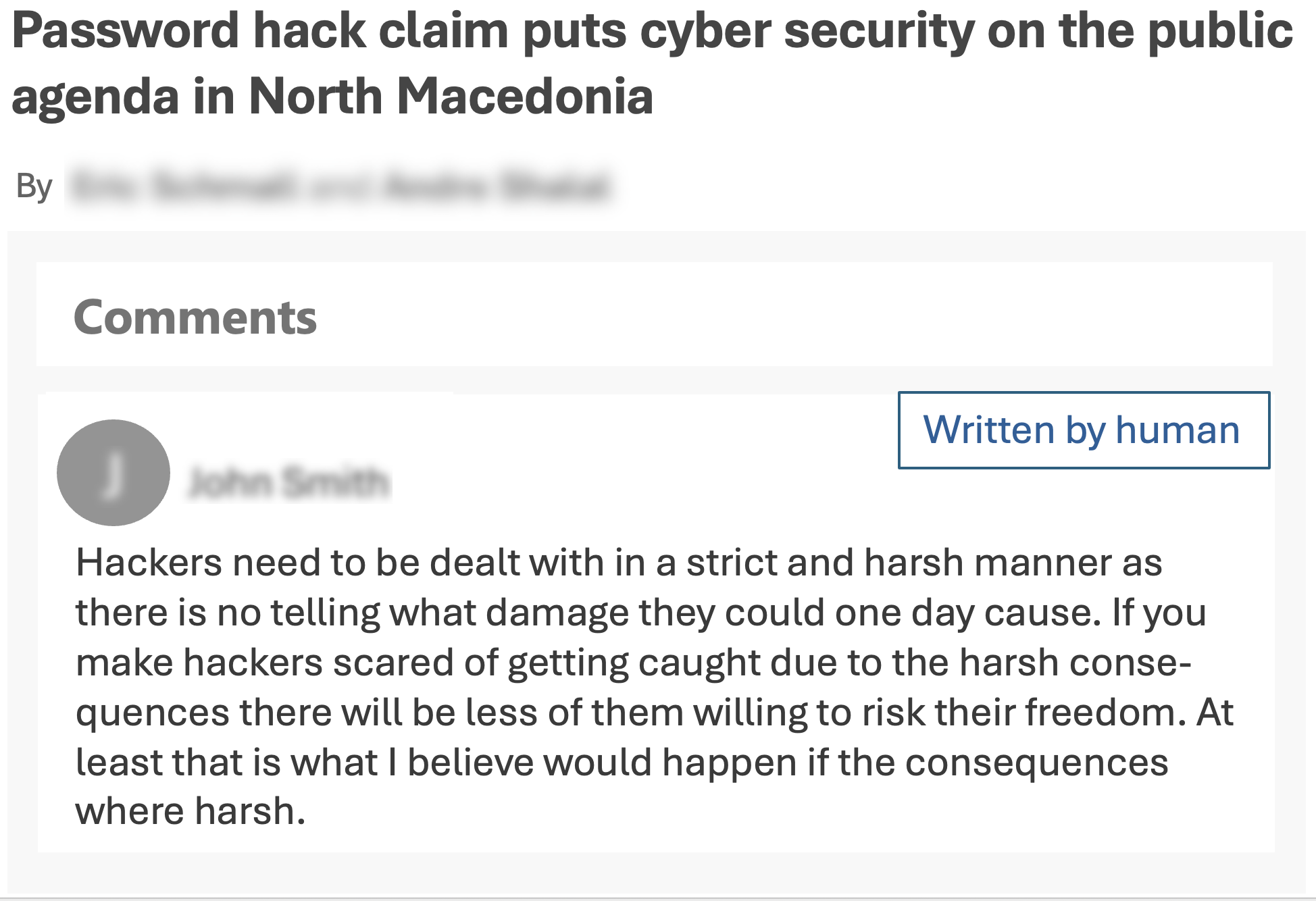}
  \caption{An example of the stimulus presentation under the source condition: \textit{written by human}.}
  \label{fig:stimuli}
\end{figure}

The selected articles were partitioned into three topic-balanced groups of 16 articles each $(A, B, C)$, which were then combined into six counterbalanced stimulus sets. Each set contained 32 articles balanced across fallacy and non-fallacy conditions. Paired sets contained identical articles but reversed the assignment of fallacious versus non-fallacious comments, ensuring that each article appeared in both conditions across the dataset. Each set includes two instances of each of the eight fallacy types. For stimuli presentation, we adopt a standardized news-style interface inspired by Reuters, which is widely rated as politically neutral by media-bias assessment organizations~\cite{adfontesmedia, allsides}. The interface is minimal, avoiding identifiable branding and blurring user names to mitigate unintentional biases; Figure \ref{fig:stimuli}. As Reuters does not support comment threads, we adapted a BBC-inspired comment layout to approximate a realistic reading experience. 





\subsection{Human Evaluation}
We conducted a mixed-design experiment with a 5-level between-subjects factor (Human, AI, Human+AI, AI+Human, and Control)\footnote{For brevity, \textbf{\textit{human with AI assistance}} and \textbf{\textit{AI with human assistance}} are denoted as \textbf{Human+AI} and \textbf{AI+Human}.} and a 2-level within-subjects factor (fallacy: present vs. absent). Participants viewed identical content under different source labels. This study was approved by the Institutional Review Board (IRB) at the authors' institution.

\subsubsection{Participants}
The experiment was designed using Qualtrics and conducted on Prolific. Participants were at least 18 years old, located in the United States, and fluent in English. Power analysis using G*Power 3.1 \cite{faul2009statistical} suggested $n = 470$ participants to detect a small effect size (Cohen’s $f = 0.1$) of the interaction of source condition and fallacy presence, with a power of $0.95$ [analysis of variances (ANOVA) test, $\alpha = .05$]. To account for potential submission removals while ensuring statistical power, we recruited $510$ Prolific participants on January 28, 2026. We used Prolific’s built-in quality controls to reject submissions that were exceptionally fast or potentially AI-generated, and accepted $505$ participants  (Human = 103, AI = 103, Human+AI = 100, AI+Human = 101, Control = 98).\footnote{Submission removals: three incomplete, two failed attention checks, none with duplicate IP or GPS coordinates \cite{nahar2024fakes}.}  Demographics were similar across conditions; Appendix \ref{appendix: participant demographic}. The median completion time was 26 minutes 10 seconds, and participants were paid \$4 (hourly rate: \$9.17; above the U.S. minimum \$7.25). Participants who failed attention checks were paid partially.



\begin{figure}[h]
\centering
  \includegraphics[width=\columnwidth]{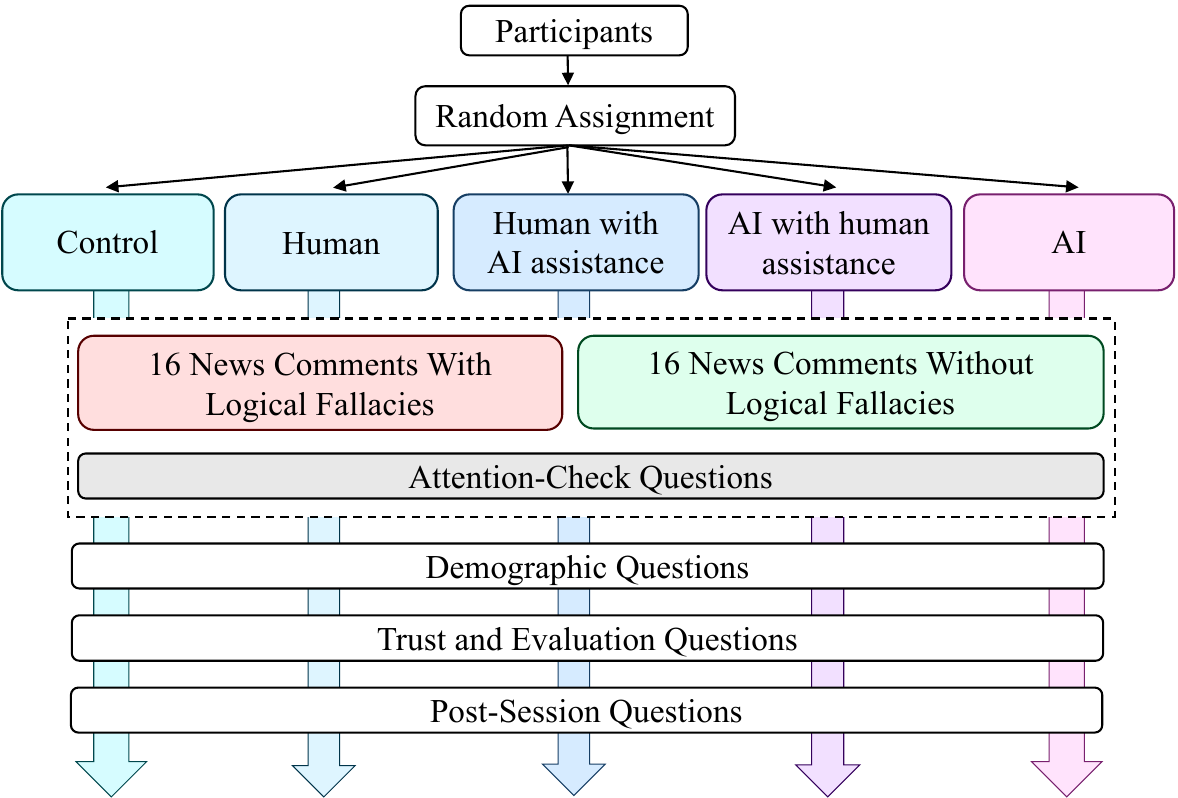}
  \caption{An overview of the human-subjects study. Participants were randomly assigned to a source condition and evaluated news comments with and without logical fallacies, with randomized comment ordering.}
  \label{fig:study_design}
\end{figure}


\subsubsection{Procedure and Measures}
Participants were randomly assigned to any of the five source conditions, and viewed one of the six news title-comment sets; Figure \ref{fig:study_design}; Appendix \ref{subsec: stimuli and presentation}. After providing informed consent and receiving a brief introduction to logical fallacies, participants evaluated 32 news title–comment pairs (Figure \ref{fig:stimuli}), balancing coverage across fallacy types. Stimuli were presented in a randomized order. For each stimulus, participants rated the logical accuracy and their confidence in their accuracy judgments on 5-point scales; Appendix \ref{appendix: procedures and measures}. They randomly encountered two attention-check questions and the survey automatically terminated for those who failed to pass either of the attention checks. 

Afterward, participants provided demographic information and reported their AI expertise, political affiliation, etc. These variables were statistically similar across conditions (all $ps > .05$) and were not included as covariates in the main analyses. However, as additional robustness analyses, we included these variables as covariates and the conclusions remained unchanged; Appendix \ref{appendix: additional robustness analyses}.

Participants completed a manipulation check and rated their trust in the commenter across affective, cognitive, and behavioral dimensions \cite{chen2023ai}, and their overall evaluation (e.g., helpfulness, competence) \cite{nahar2025catch}. Both constructs were measured using multi-item 5-point Likert scales. The trust (Cronbach’s $\alpha = .84$) and evaluation ($\alpha = .89$) measures demonstrated good internal consistency, and responses were averaged to create composite scores for subsequent analyses. Participants also ranked the sources and described how they thought the comments were written. Finally, they were debriefed that all comments were written by humans with AI assistance. Demographic and post-session details are in Appendix \ref{appendix: participant demographic}, \ref{appendix: procedures and measures}, \ref{appendix: post_session_results}.

\subsection{LLM Evaluation}
To enable a direct comparison with human judgments, we evaluate three LLMs: GPT-5.2 \citep{singh2025gpt5}, Gemini 2.5 Flash \cite{comanici2025gemini2.5}, and Claude Sonnet 4.5 \citep{anthropic2025claudesonnet45} on the same set of comments. As human participants received instructions regarding the source labels at the beginning of the study, the LLM prompts included the same instructions to ensure that both humans and LLMs paid attention to the source labels. The stimuli and rating scales were also kept identical to those presented to human evaluators (Appendix \ref{appendix: method_llm_evaluation}), with models rating perceived logical accuracy and confidence on 5-point scales. Please refer to Appendix \ref{appendix: robust analysis across prompting strategies} for robustness analysis across prompting strategies: chain-of-thought prompting \cite{wei2022chain} and expert-evaluator framing \cite{xu2023expertprompting}, and \ref{appendix: additional robustness analyses} for robustness across open-weight models Qwen-3.5-9B \cite{qwen3.5} and Gemma-4-12B \cite{gemmateam2026gemma4technicalreport} with self-reported and calibrated confidence probabilities. 

We used the default API settings (temperature=0) for reproducibility and evaluated each stimulus independently to avoid context carryover across inputs. Following prior LLM-as-judge work \citep{gu2026survey, pan2024human}, we repeated each evaluation three times to account for potential variability in API-based model outputs. As trust in the commenter and overall evaluation reflect interpersonal social judgments most meaningful to humans, these measures are not collected in LLM evaluation.

\section{Results}
We used linear mixed-effects regression (LMER) models to better account for variability in human evaluation settings compared to traditional methods \cite{howcroft2021happens}. The LMER models included participant, item, and stimulus set as random intercepts to account for repeated observations and stimulus-level variability; Appendix \ref{appendix: data analysis}. We report ANOVA results with Satterthwaite-approximated degrees of freedom \cite{kuznetsova2017lmertest}, effect sizes using $\eta_p^2$ \cite{lakens2013calculating}, and pairwise comparisons with Tukey correction \cite{schuff2021does}. Post-session analyses are reported in Appendix \ref{appendix: post_session_results}. As additional robustness analyses, we fit cumulative link mixed models (CLMMs) for the ordinal dependent variables perceived logical accuracy and confidence, and the substantive conclusions remained unchanged; Appendix \ref{appendix: additional robustness analyses}. 

\subsection{Effect of Source on Perceived Logical Accuracy (RQ1a)}
\begin{figure}[t]
\centering
  \includegraphics[width=\columnwidth]{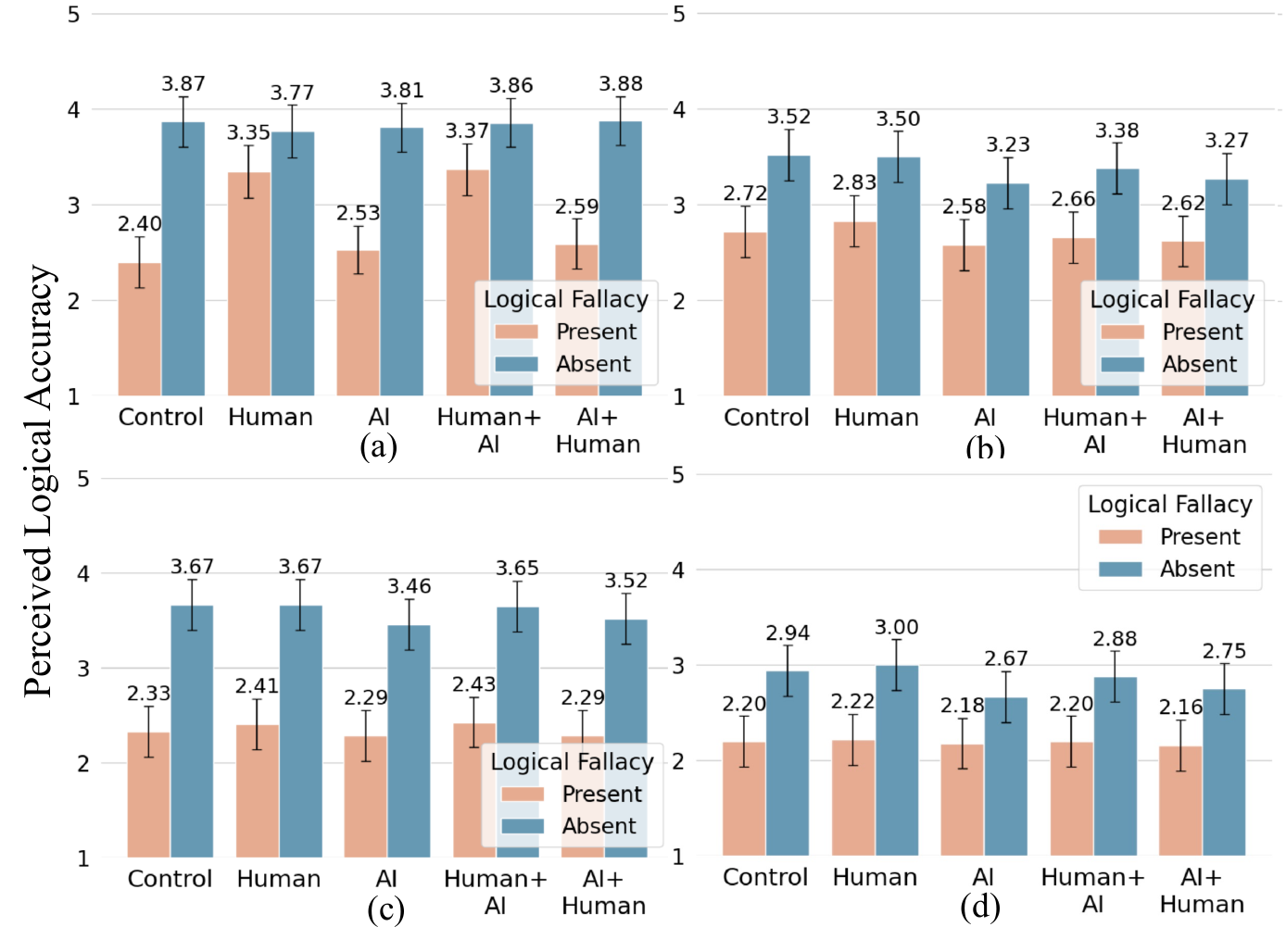}
  \caption{Average values of perceived logical accuracy ratings for (a) human participants (b) GPT-5.2 (c) Gemini 2.5 Flash (d) Claude Sonnet 4.5. Error bars represent the 95\% confidence intervals of the mean.
  }
  \label{fig:perceived_logical_accuracy_results}
\end{figure}

\textbf{Human participants were highly susceptible to logical fallacies in the Human and Human+AI conditions compared to all other conditions}. Perceived logical accuracy was strongly influenced by both source condition and the presence of logical fallacies, with a significant Condition $\times$ Fallacy interaction ($p < .001$; Table~\ref{tab:rq1a_results}). Across all source conditions, participants rated comments containing logical fallacies as significantly less logically accurate than non-fallacious comments ($p < .001$; Figure~\ref{fig:perceived_logical_accuracy_results}; Appendix \ref{appendix: pa_descriptive_stat}). 
However, this penalty for fallacies was substantially smaller in the Human ($M = 3.35$) and Human+AI ($M = 3.37$) conditions, indicating a smaller gap between fallacies and non-fallacies compared with the control, AI, and AI+Human conditions ($Ms =$ 2.40, 2.53, and 2.59; Appendix \ref{appendix: pa_fallacy_penalty}). This relationship remained significant after restricting the analysis to participants who correctly identified their assigned source labels; Appendix \ref{appendix: manipulation-check analysis}. 
In contrast, non-fallacy comments were rated similarly across all source conditions, suggesting that source labels only influenced judgments when reasoning quality was poor. Overall, participants in the Human and Human+AI conditions ($M =$ 25 min 5 sec) spent less time evaluating the comments compared to the other conditions ($M=$ 28 min 17 sec, $p < .01$), suggesting greater reliance on heuristic rather than deliberative evaluation in those conditions.

\begin{table}[h]
\centering
\small
\begin{tabular}{lcc}
\toprule
\textbf{Effect} & \textbf{Human}($F, \eta_p^2$) & \textbf{LLM} ($F, \eta_p^2$) \\
\midrule
Cond. & \textbf{17.84, .12$^{***}$} & \textbf{7.16, .02$^{***}$} \\
FL & \textbf{1943.84, .80$^{***}$} &  \textbf{41.71, .14$^{***}$} \\
Model & -- & \textbf{123.59, .16$^{***}$} \\
Cond. $\times$ FL & \textbf{97.14, .44$^{***}$} & 1.22, $<.01$ \\
Cond. $\times$ Model & -- & 0.34, $<.01$ \\
FL $\times$ Model & -- & \textbf{43.04, .06$^{***}$} \\
Cond. $\times$ FL $\times$ Model & -- & 0.23, $<.01$ \\
\bottomrule
\end{tabular}
\caption{ANOVA results for perceived logical accuracy. Note. \lq\lq Cond.\rq\rq = \lq\lq Condition\rq\rq, \lq\lq FL\rq\rq = \lq\lq Fallacy\rq\rq. $^{***}p<.001$.}
\label{tab:rq1a_results}
\end{table}

\textbf{Unlike humans, LLMs were largely stable across conditions}. While there was a main effect of condition, there were no significant pairwise differences after correcting for multiple comparisons. However, perceived logical accuracy showed strong main effects of both model and fallacy presence. Across conditions, comments containing logical fallacies were rated as significantly less logically accurate than non-fallacious comments ($p < .01$). While LLMs were resistant to source-label bias, they differed significantly in evaluative strictness and sensitivity to flawed reasoning. For non-fallacy comments, Gemini 2.5 Flash ($M = 3.59$) assigned the highest ratings, GPT-5.2 ($M = 3.38$) was intermediate, and Claude Sonnet 4.5 ($M = 2.85$) was lowest. For fallacies, GPT-5.2 ($M = 2.68$) gave the highest ratings, while Gemini 2.5 Flash ($M = 2.35$) and Claude Sonnet 4.5 ($M = 2.19$) assigned lower ratings. \textbf{Gemini 2.5 Flash showed the greatest sensitivity to logical quality}, producing the largest separation between fallacious and non-fallacious comments; Appendix \ref{appendix: pa_descriptive_stat}.

LLMs judged fallacies as less logically accurate than humans did ($p <.001$), although this effect is driven primarily by human susceptibility to fallacies in the Human and Human+AI conditions. For non-fallacies, the pattern was more mixed; humans ($M = 3.84$) rated comments higher than GPT-5.2 ($M = 3.38$, $p < .001$) and Claude Sonnet 4.5 ($M = 2.85$, $p<.001$), but did not differ significantly from Gemini 2.5 Flash ($M = 3.59$). These results suggest that LLMs were generally stricter than humans, especially for fallacies. Gemini 2.5 Flash most closely matched human ratings for logically sound comments while still penalizing fallacies more strongly; Appendix \ref{appendix_pa_pairwise}.

\textbf{Fallacy-Level Results:} Both humans and LLMs varied in perceived logical accuracy across fallacy types, suggesting that some reasoning errors were more difficult to detect than others. Humans assigned  higher ratings to \textit{hasty generalization} and \textit{false dilemma} compared to several other fallacy types, while \textit{slippery slope} and \textit{appeal to majority} generally received lower ratings ($p_{adj}<.001$). In contrast, LLMs assigned higher ratings to \textit{appeal to nature} and \textit{appeal to tradition} ($p_{adj}<.001$), whereas \textit{appeal to worse problems}, \textit{false dilemma}, and \textit{slippery slope} received lower ratings. Interestingly, human and LLM error patterns suggest complementary strengths, highlighting the value of human–LLM collaboration for logical reasoning tasks; Appendix \ref{appendix:pa_fallacy_type_analysis}.

\subsection{Effect of Source on Confidence Ratings (RQ1b)}

\begin{figure}[t]
\centering
  \includegraphics[width=\columnwidth]{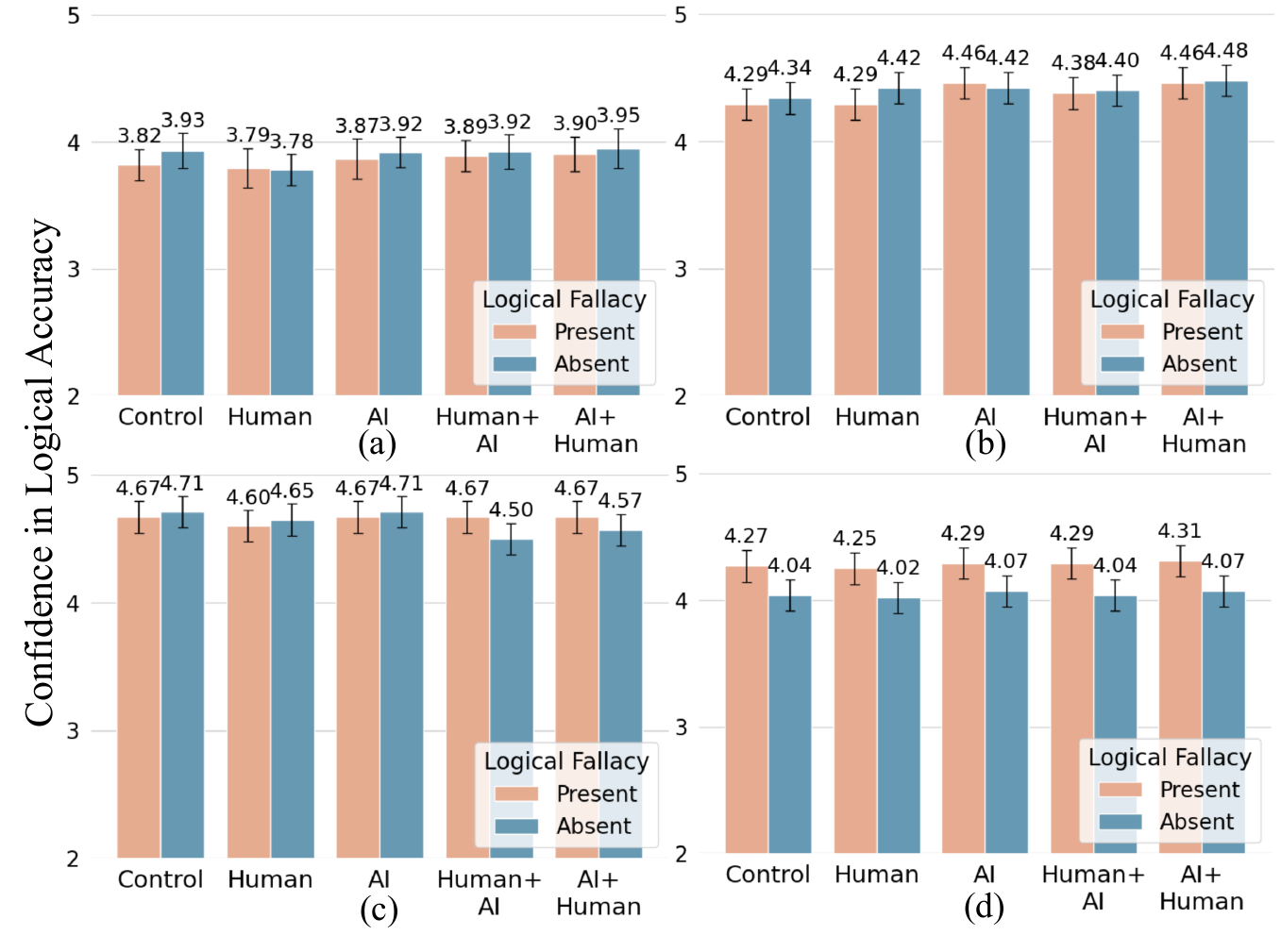}
  \caption{Average values of confidence ratings for (a) human participants (b) GPT-5.2 (c) Gemini 2.5 Flash (d) Claude Sonnet 4.5. Error bars represent the 95\% confidence intervals of the mean.
  }
  \label{fig:confidence_results}
\end{figure}

\textbf{Human participants were similarly confident in their judgments for all labeled source conditions} (Table~\ref{tab:rq1b_results}; Figure~\ref{fig:confidence_results}). Confidence was lower for comments containing logical fallacies than for non-fallacious comments ($p=.001$; Appendix \ref{appendix: confidence_descriptive_statistics}, \ref{appendix: confidence_pairwise_comparisons}), driven primarily by the control condition, where participants were less confident when evaluating fallacy ($M = 3.82$) vs. non-fallacy ($M = 3.93$, $p<.001$); differences were not significant within labeled source conditions, suggesting that source labels may inflate perceived confidence independently of actual reasoning quality. This pattern persisted when the analysis was restricted to participants who correctly identified their assigned source labels; Appendix \ref{appendix: manipulation-check analysis}.

\begin{table}[h]
\centering
\small
\begin{tabular}{lcc}
\toprule
\textbf{Effect} & \textbf{Human}($F, \eta_p^2$) & \textbf{LLM} ($F, \eta_p^2$) \\
\midrule
Cond. & .81, $<.01$ & 1.95, $<.01$ \\
FL & \textbf{10.42, .02$^{***}$} & 1.86, .01 \\
Model & -- & \textbf{229.48$^{***}$, .26} \\
Cond. $\times$ FL & 2.05, .02 & 1.24, $<.01$ \\
Cond. $\times$ Model & -- & 1.66, .01 \\
FL $\times$ Model & -- & \textbf{20.44$^{***}$, .03} \\
Cond. $\times$ FL $\times$ Model & -- & 0.76, $<.01$ \\
\bottomrule
\end{tabular}
\caption{ANOVA results for confidence ratings. Note. \lq\lq Cond.\rq\rq = \lq\lq Condition\rq\rq, \lq\lq FL\rq\rq = \lq\lq Fallacy\rq\rq. $^{***}p<.001$.}
\label{tab:rq1b_results}
\end{table}

\textbf{LLMs were also stable across source labels}, other than two small pairwise differences for Gemini 2.5 Flash for non-fallacy, where Control and AI showed slightly higher confidence ratings than the Human+AI condition ($p = .025$). However, substantial differences emerged across models. Gemini 2.5 Flash assigned higher confidence scores, GPT-5.2 was intermediate, and Claude Sonnet 4.5 was the lowest. Claude Sonnet 4.5 was more confident for fallacies, whereas others showed little change; Appendix \ref{appendix: confidence_pairwise_comparisons}.

LLM confidence judgments were less conservative than human judgments. All three LLMs assigned higher confidence scores for both fallacy and non-fallacy comments compared to humans ($p<.001$), with model means ranging from approximately 4.0 to 4.7 on a 1-5 scale, compared to human confidence ratings (fallacy: 3.86; non-fallacy: 3.90); Appendix \ref{appendix: confidence_pairwise_comparisons}. 

\textbf{Fallacy-Level Results:} Both humans and LLMs varied in confidence ratings across fallacy types. Humans reported higher confidence for \textit{slippery slope}, \textit{appeal to worse problems}, and \textit{appeal to majority} ($p_{adj}<.01$), whereas confidence was comparatively lower for \textit{hasty generalization}. For LLMs, confidence was higher for \textit{appeal to tradition} and \textit{appeal to nature} ($p_{adj} < .001$), while \textit{appeal to worse problems} and \textit{false dilemma} had lower confidence ratings; Appendix \ref{appendix: confidence_fallacy_type_analysis}. Humans and LLMs exhibited different confidence-susceptibility patterns across fallacy types: humans were sometimes more susceptible despite lower confidence, whereas LLM confidence more closely aligned with perceived logical accuracy judgments.

\subsection{Effect of Source on Trust and Overall Evaluation (RQ2)}
\textbf{The Human and Human+AI conditions received significantly higher trust and overall evaluation ratings.}
(Table~\ref{tab:rq2_results}). For trust, Human exceeded Control, AI, and AI+Human (all $ps<.001$), while Human+AI exceeded Control ($p<.001$) and AI ($p=.003$). Overall evaluation followed a similar pattern: Human exceeded Control, AI, and AI+Human (all $ps<.001$), while Human+AI exceeded Control ($p<.001$), AI ($p<.001$), and AI+Human ($p=.006$); Appendix \ref{appendix: trust and evaluation results}. These results mirror the perceived logical accuracy ratings for the fallacy condition. Additionally, trust and evaluation scores were strongly correlated ($r=.89, p<.001$), i.e., participants who trusted the commenter more also evaluated the commenter more positively.

The results for trust remained unchanged after restricting the analysis to participants who correctly identified their assigned source label ($N=415$). However, for overall evaluation, the Human+AI advantage over AI+Human in overall evaluation was only suggestive ($p=.072$). Nevertheless, the evaluation ratings for the Human condition remained significantly higher than Control, AI, and AI+Human (all $ps<.001$); Appendix \ref{appendix: manipulation-check analysis}.

\begin{table}[h]
\centering
\small
\begin{tabular}{lcc}
\toprule
\textbf{Effect/Source} & \textbf{Trust} & \textbf{Evaluation} \\
\midrule

Condition, $F(4,500)$ & \textbf{$16.00^{***}$ } & $17.75^{***}$ \\
Effect size, $\eta^2_p$ & $.11$ & $.12$\\
\midrule
Control & 2.83 & 3.11 \\
Human & \textbf{3.69} & \textbf{3.91} \\
AI & 2.93 & 3.15 \\
Human+AI & \textbf{3.39} & \textbf{3.63} \\
AI+Human & 3.06 & 3.23 \\
\bottomrule
\end{tabular}
\caption{ANOVA results and means for trust and overall evaluation ratings. $^{***}p<.001$. Bold fonts indicate statistically significant pairwise differences.}
\label{tab:rq2_results}
\end{table}

\section{Discussion}

\paragraph{Human reasoning evaluation is sensitive to source labels.} Participants were more susceptible to logical fallacies labeled as human or human with AI assistance and assigned higher trust ratings in these conditions, despite identical comment content and comparable demographics and AI expertise across conditions. According to the heuristic-systematic model \cite{chen1999heuristic}, people rely more on heuristic cues when tasks are cognitively demanding or when motivation and cognitive resources are limited, leading them to make quicker judgments based on cues such as source attractiveness or credibility \cite{o2013elaboration, pennycook2021psychology, todorov2002heuristic}. Logical fallacies are difficult to detect: even expert annotators exhibit substantial disagreement on the CoCoLoFa dataset \cite{yeh2024cocolofa}. Indeed, participants spent less time evaluating comments in the Human and Human+AI conditions, suggesting shallower, heuristic-driven processing. Paradoxically, participants ranked Human and Human+AI as their most preferred source conditions (Appendix \ref{appendix_post-session_source_preferences}), yet these were precisely the conditions most associated with susceptibility to fallacies, suggesting that the cues participants found the most reassuring were also the most misleading.

\paragraph{Hybrid authorship labels shape perceived credibility.} It is perhaps unsurprising that human-written content is perceived more favorably \cite{rae2024effects}. However, reliance on AI-assisted systems has grown rapidly, reaching 53\% population adoption within three years, faster than the PC or the internet \cite{stanfordaiindex}. As hybrid authorship becomes increasingly common, these labels may influence how people evaluate information quality and credibility. An exploratory analysis revealed that \textit{participants interpreted human with AI assistance and AI with human assistance as different forms of authorship}: the former was perceived as human-originated ideas polished by AI, whereas the latter was perceived as AI-originated ideas refined by humans; Appendix \ref{appendix_post-session_comment_writing_process}. This asymmetry likely explains the elevated trust and susceptibility observed in the Human+AI condition. When human authorship is perceived as primary, content may inherit the credibility people typically associate with human-originated ideas. Thus, malicious actors may exploit human-involvement cues to lend unwarranted credibility to logically flawed arguments, making them more persuasive than they should be.

\paragraph{Implications for human-LLM reasoning workflows.} LLMs were comparatively stable across source conditions when assessing reasoning quality. While prior work reports label effects in trust-oriented evaluations \cite{sun2026label}, we found LLM judgments of logical fallacies to be largely robust to source manipulations, suggesting that source sensitivity may be task and model-dependent rather than universal. 
This could also be explained by recent advances that have improved LLM performance on reasoning tasks~\cite{jeong2025large}, though LLMs remain prone to their own systematic biases even when producing structured and coherent judgments~\cite{zheng2023judging}.

Still, models differed meaningfully in evaluation style and sensitivity to logical quality. Gemini 2.5 Flash showed the strongest separation between fallacies and non-fallacies, whereas Claude Sonnet 4.5 produced more conservative ratings overall. Importantly, the source-agnostic evaluation patterns were robust across prompting strategies; Appendix \ref{appendix: robust analysis across prompting strategies}. These findings suggest that decision-making workflows may benefit from carefully designed human–LLM collaboration pipelines to help reduce vulnerabilities unique to either humans or LLMs.

\paragraph{Humans and LLMs are more susceptible to different types of logical fallacies.} Logical fallacy detection remains a challenging problem \cite{jin2022logical}. Even expert human annotators exhibit substantial disagreement on the CoCoLoFa dataset \cite{yeh2024cocolofa}, and both humans and LLMs are more susceptible to certain fallacy types than others \cite{payandeh2024susceptible}. Our findings revealed a similar pattern: humans were more susceptible to \textit{hasty generalizations}, whereas LLMs were more influenced by \textit{appeal to nature} \textit{and appeal to tradition} fallacies. Thus, humans and LLMs may rely on distinct reasoning heuristics when evaluating argument quality, leading to complementary error patterns, supporting the design of collaborative workflows that use LLMs not as replacements for human judgment, but as complementary checks against source-driven human biases.

\paragraph{Confidence judgments do not necessarily reflect reasoning quality.} Humans remained similarly confident across labeled source conditions despite being more susceptible to fallacies in the Human and Human+AI conditions, suggesting a mismatch between confidence and reasoning quality. However, they were somewhat less confident in assessing fallacies in the control condition, where no source label was provided, suggesting that source labels may serve as heuristic cues that increase perceived confidence even when they do not improve reasoning quality. LLM confidence ratings were also stable across source conditions, but LLMs assigned higher confidence scores than humans, even for fallacious content, raising concerns about overconfidence. Confidence alone may not reliably reflect reasoning quality for either humans or LLMs \cite{sun2025large}. Moreover, fallacy-type analyses suggest that humans and LLMs differ in confidence calibration: humans were sometimes more susceptible to certain fallacy types despite lower confidence, whereas LLM confidence more closely aligned with perceived logical accuracy judgments; Appendix \ref{appendix: human_and_llms_differ_in_confidence_calibration}.

\paragraph{Design implications for increasingly AI-mediated online systems.} Source disclosures do not simply inform users about where content comes from; they may unintentionally encourage heuristic judgments over deliberate evaluation of reasoning quality. This is especially consequential given that our participants primarily consume news through social media (Appendix \ref{appendix_post-session_news_source}), where source labels could be easily manipulated by malicious actors. Yet simply removing source labels is not a viable solution: no-disclosure condition was the least preferred, suggesting that absent labels may not be well-tolerated by users without necessarily improving reasoning quality; Appendix \ref{appendix_post-session_source_preferences}. Instead, our findings highlight the need for systems that support content-focused evaluation, such as automated detection of reasoning errors, contextual warnings, or interfaces that encourage users to assess arguments on logical quality rather than source cues alone \cite{kumar2024watch, pennycook2020fighting}.

\section{Conclusion}
As AI-generated and AI-assisted content becomes increasingly prevalent, it is critical to understand how source perception shapes human judgment. We show that humans are susceptible to source-label bias when assessing logical reasoning: content labeled as written by a human or human with AI assistance receives higher trust ratings and is more likely to be judged as logically accurate despite containing fallacies. Thus, perceived source has a significant impact on human reasoning evaluation. While LLMs are largely source-agnostic compared to humans, their performance remains model-dependent, indicating that they are not a standalone solution. Instead, carefully designed human–LLM collaboration pipelines may help mitigate vulnerabilities unique to either humans or LLMs alone.

\section*{Limitations}
Our study has several limitations. First, we specifically focus on logical fallacies in short online comments using the CoCoLoFa dataset \cite{yeh2024cocolofa}, and our conclusion applies to the logical fallacy evaluation for the specific setting studied here, rather than to all forms of source-label bias or all LLM evaluation contexts. While logical fallacies provide a controlled way to evaluate reasoning quality independently of domain-specific factual knowledge, it might be worthwhile to study additional contextual, social, or multimodal factors in future work. 

Second, our source conditions were intentionally simplified and limited to a small set of disclosure formats. While we examined logical fallacies where source labels do not carry useful information, real online platforms may contain more complex and ambiguous signals of authorship and credibility. Thus, future work could examine cases where source labels may provide useful credibility signals and test whether humans and LLMs can incorporate valid source information without over-relying on it. This is especially important because source labels can also be manipulated by malicious actors, even in tasks where source information is otherwise useful. While it is tempting to incorporate multiple datasets, more nuanced fallacy types, and additional factors influencing source-label presentation in a single study, human-subject studies impose practical trade-offs: increasing experimental complexity may make it harder to isolate effects and reduce statistical power.

Third, our human-subjects experiment was conducted using Prolific workers from the United States, who are typically English-speaking, educated, and technologically aware \cite{douglas2023data}, which might impact the generalizability of our findings. Moreover, participants were exposed to a relatively high proportion of logical fallacies (16 out of 32), which presumably exceeds the likelihood of encountering logical fallacies in real-world settings. While this was to ensure sufficient variation across logical fallacy types (two examples from each of the eight fallacy categories), future research should examine how reduced frequency of logical fallacies might alter the current findings, thereby assessing their ecological validity. Importantly, our analyses found no significant main or interaction effects of presentation order, suggesting that randomizing question order helped minimize potential order-related biases.

Fourth, our LLM evaluation results are limited to three contemporary models (GPT-5.2, Gemini 2.5 Flash, and Claude Sonnet 4.5) and different prompting strategies, including the robustness analyses. While we found LLMs to be comparatively stable across source conditions for our prompting strategies, this should not be interpreted as evidence that they are universally unbiased or reliable evaluators, as such evaluations depend on the domain and task.  Notably, LLM evaluations reflect model-generated outputs rather than underlying reasoning processes and may be sensitive to prompt formulation, model choice, and future system updates.

\section*{Ethical Considerations}
All procedures were conducted in accordance with relevant laws and institutional guidelines. Participants’ privacy rights were protected, and informed consent was obtained prior to the study. The study protocol was approved by the Institutional Review Board (IRB) at the authors’ institution, The Pennsylvania State University (IRB reference: STUDY00028127, approval date: November 25, 2025). Although participants were informed that the study involved logical fallacies, the experiment employed deception regarding source disclosure to preserve ecological validity. Following IRB guidelines, participants were debriefed at the end of the study, including an explanation of the withheld information and access to appropriate support resources if needed.

Our findings also raise broader ethical concerns regarding AI-mediated communication. In particular, participants showed increased trust and susceptibility when content was labeled as written by a human or written by a human with AI assistance, suggesting that logically erroneous content paired with signals of human involvement may appear more credible and persuasive. Such effects could potentially be exploited to amplify misleading or manipulative content at scale, especially since source labels can be easily misrepresented online. At the same time, studying these phenomena in controlled research settings is important for understanding how source cues influence reasoning judgments and for informing the design of safer and more robust online systems.

\section*{Acknowledgments}
This work was supported in part by U.S. NSF awards no. 2438810 and 2555559, 2026 Amazon Nova AI challenge award, and PSU Frymoyer chairship. Some experimental results were obtained using computational resources provided by CloudBank, supported through U.S. NAIRR award no. 240336. In addition, we acknowledge the support from the  Linguistic Diversity Across the Lifespan Graduate Research Traineeship Program (NSF grant no. 2125865).

\bibliography{custom}

\appendix
\section{Appendix}
\label{sec:appendix}

\subsection{AI Assistance Disclosure}
We acknowledge the use of AI tools in supporting roles throughout this paper. AI tools assisted with tasks such as improving writing clarity and grammar, understanding technical concepts, and producing initial drafts of code, figures, and tables. The authors thoroughly reviewed, validated, and revised all AI-assisted materials before use. Notably, all conceptual ideas, study design decisions, analyses, and interpretations were developed by the authors. AI-generated outputs served only as preliminary assistance and were subject to substantial human oversight and verification. The authors assume full responsibility for the accuracy, integrity, and originality of the work.

\subsection{Fallacy Types} \label{appendix: fallacy types}
The definitions, examples, and explanations of the eight logical fallacy types used in this paper are drawn directly from the CoCoLoFa dataset \cite{yeh2024cocolofa} and Logically Fallacious\footnote{https://www.logicallyfallacious.com/}.

\textbf{Appeal to authority.} \textit{Definition:} Insisting that \textbf{a claim is true simply because a valid authority or expert on the issue said it was true}, without any other supporting evidence offered. \textit{Example:} Richard Dawkins, an evolutionary biologist and perhaps the foremost expert in the field, says that evolution is true. Therefore, it's true. \textit{Explanation:} Richard Dawkins certainly knows about evolution, and he can confidently tell us that it is true, but that doesn't make it true. What makes it true is the preponderance of evidence for the theory.

\textbf{Appeal to majority.} \textit{Definition:} \textbf{When the claim
that most or many people in general or of a particular group accept a belief as true is presented as evidence for the claim}. Accepting another person’s belief, or many people’s beliefs, without demanding evidence as to why that person accepts
the belief, is lazy thinking and a dangerous way
to accept information. \textit{Example:} Up until the late 16th century, most people believed that the earth was the center of the universe.  This was seen as enough of a reason back then to accept this as true. \textit{Explanation:} The geocentric model was an observation (limited) and faith-based, but most who accepted the model did so based on the common and accepted belief of the time, not on their own observations, calculations, and/or reasoning.

\textbf{Appeal to nature.} \textit{Definition:} When used as a fallacy, \textbf{the belief or suggestion that natural is better than unnatural based on its naturalness}. Many people adopt this as a default belief. It is the belief that is what is natural must be good (or any other positive, evaluative judgment) and that which is unnatural must be bad (or any other negative, evaluative judgment). \textit{Example:} Cocaine is all natural; therefore, it is good for you. \textit{Explanation:} There are very many things in this world that are natural and very bad for you besides cocaine, including, earthquakes, monsoons, and viruses, just to name a few.  Whereas unnatural things such as aspirin, pacemakers, and surgery can be very good things.

\textbf{Appeal to tradition.} \textit{Definition:} \textbf{Using historical preferences of the people (tradition), either in general or as specific as the historical preferences of a single individual, as evidence that the historical preference is correct}.  Traditions are often passed from generation to generation with no other explanation besides, \lq\lq this is the way it has always been done\rq\rq. This is not a reason; it is an absence of a reason. \textit{Example:} Dave: For five generations, the men in our family went to Stanford and became doctors, while the women got married and raised children.  Therefore, it is my duty to become a doctor. Kaitlin: Do you want to become a doctor? Dave: It doesn’t matter -- it is our family tradition.  Who am I to break it? \textit{Explanation:} Just as it takes people to start traditions, it takes people to end them.  A tradition is not a reason for action -- it is like watching the same movie over and over again but never asking why you should keep watching it.

\textbf{Appeal to worse problems.} \textit{Definition:}  \textbf{Trying to make a scenario appear better or worse by comparing it to the best or worst case scenario. }\textit{Example:} Son: I am so excited!  I got an \lq\lq A\rq\rq \space on my physics exam! Dad:  Why not an \lq\lq A+\rq\rq?  This means that you answered something incorrectly. That is not acceptable! \textit{Explanation:} The poor kid is viewing his success from a very reasonable perspective based on norms.  However, the father is using a best case scenario as a comparison, or a very unreasonable perspective.  The conclusion \lq\lq it is not acceptable,\rq\rq \space is unreasonable and, therefore, fallacious.

\textbf{False dilemma.} \textit{Definition:} \textbf{When only two choices are presented yet more exist, or a spectrum of possible choices exists between two extremes.}  False dilemmas are usually characterized by \lq\lq either this or that\rq\rq \space language, but can also be characterized by omissions of choices.  \textit{Example:} I thought you were a good person, but you weren’t at church today. \textit{Explanation:} The assumption here is that if one doesn't attend chuch, one must be bad.  Of course, good people exist who don’t go to church, and good church-going people could have had a really good reason not to be in church.

\textbf{Hasty generalization.} \textit{Definition:} \textbf{Drawing a conclusion based on a small sample size, rather than looking at statistics that are much more in line with the typical or average situation.} \textit{Example:} Four out of five dentists recommend Happy Glossy Smiley toothpaste brand.  Therefore, it must be great. \textit{Explanation:} It turns out that only five dentists were actually asked.  When a random sampling of 1000 dentists was polled, only 20\% actually recommended the brand.  The four out of five result was not necessarily a biased sample or a dishonest survey; it just happened to be a statistical anomaly common among small samples.

\textbf{Slippery slope.} \textit{Definition:} \textbf{When a relatively insignificant first event is suggested to lead to a more significant event, which in turn leads to a more significant event, and so on, until some ultimate, significant event is reached, where the connection of each event is not only unwarranted but with each step it becomes more and more improbable}.  Many events are usually present in this fallacy, but only two are actually required -- usually connected by \lq\lq the next thing you know...\rq\rq \textit{Example:} We cannot unlock our child from the closet because if we do, she will want to roam the house.  If we let her roam the house, she will want to roam the neighborhood.  If she roams the neighborhood, she will get picked up by a stranger in a van, who will sell her in a sex slavery ring in some other country.  Therefore, we should keep her locked up in the closet. \textit{Explanation:} In this example, it starts out with reasonable effects to the causes.  For example, yes, if the child is allowed to go free in her room, she would most likely want to roam the house.  Sure, if she roams the house, she will probably want the freedom of going outside, but not necessarily roaming the neighborhood.  Now we start to get very improbable.  The chances of her getting picked up by a stranger in a van to sell her into sex slavery in another country is next to nothing.

\subsubsection{Selected Topics and Filtering Strategy} \label{selected topics and filtering strategy}
CoCoLoFa uses topic tags such as protest, international relations, race issue, women rights, Russo-Ukrainian war, environmental issue, gender issue, human rights, drug issue, police brutality, immigration/refugees, COVID/health issue, legislation, freedom of speech, election, sustainability, religious conflict, political debates, U.S. politics, digital rights, and East Asian politics, each associated with the top 10 keywords. However, instead of directly relying on these tags, we adopted a keyword-based topic assignment approach because the provided labels did not always align well with the actual article content. For example, some articles categorized under \textit{culture} were mapped to \textit{race issue}, although such mappings did not always correspond well to the intended topic distinctions in the U.S. context.

\begin{table}[h]
\centering
\small
\begin{tabular}{lccc}
\toprule
\textbf{Topic} & \textbf{Group 1} & \textbf{Group 2} & \textbf{Group 3} \\
\midrule
Digital rights        & 6 & 6 & 6 \\
Environmental issue   & 4 & 4 & 4 \\
Culture                & 3 & 3 & 3 \\
Sustainability         & 1 & 0 & 1 \\
Economic issue         & 0 & 1 & 1 \\
Traffic                & 1 & 1 & 0 \\
Health                 & 0 & 1 & 1 \\
Public trust           & 1 & 0 & 0 \\
\midrule
\textbf{Total}         & \textbf{16} & \textbf{16} & \textbf{16} \\
\bottomrule
\end{tabular}
\caption{Topic distribution of the stimuli used.}
\label{tab:topic_distribution}
\end{table}

As we focused on topics that were not politically biased or emotionally charged, we conducted a two-step filtering process: manual and LLM-based. First, news topics were manually screened to exclude content related to geopolitical tensions, ongoing conflicts, and other sensitive domains.  Then, we employed LLM-based filtering with Claude Sonnet 4.5 and GPT-4, using the following prompt: \textit{\lq\lq Analyze the given news title based on whether it is politically biased and emotionally charged in the context of the United States: [News title] and score it on a scale of 1 to 5 (1= not at all politically biased or emotionally charged, 2 = only slightly politically biased or emotionally charged, 3 = somewhat politically biased or emotionally charged, 4 = moderately politically biased and emotionally charged, 5 = completely politically biased and emotionally charged) \rq\rq}. We only selected the news titles that were classified as \lq\lq 1\rq\rq \space or \lq\lq 2\rq\rq \space by both models. After the two-step filtering process (manual and LLM-based), we were left with 48 news titles on the following topics: culture, digital rights, environmental issue, sustainability, traffic, health, economic issue, and public trust. The news articles are divided into three groups of 16 articles each, with topics balanced across groups (Table \ref{tab:topic_distribution}).

\subsection{Human Evaluation}

\subsubsection{Participant Demographic} \label{appendix: participant demographic}

\textbf{Demographic Questions.} Participants answered the following demographic questions. 
\begin{itemize}
    \item Which age range do you fall into? \textit{Answer options.} 18-29 years old, 30-39 years old, 40-49 years old, 50-59 years old, 60-69 years old, 70-79 years old, 80 years or older, Prefer not to answer
    \item What is your gender? \textit{Answer options.} Male, Female, Non-Binary, Prefer not to answer
    \item Are you a native English speaker? \textit{Answer options.} Yes, No
    \item If participants answered \textit{No} to the previous question: How will you rate your English proficiency on a 5-point scale? Here, 1 means elementary proficiency and 5 means full-bilingual proficiency. \textit{Answer options.} Elementary proficiency (1), Limited working proficiency (2), Professional working proficiency (3), Full professional proficiency (4), Full bilingual proficiency (5)
    \item What is your ethnic/racial category? (You can choose the closest one.) \textit{Answer options.}  American Indian or Alaska Native, Asian, Black or African American, Hispanic or Latino, White or Caucasian, Other, Prefer not to answer
    \item What is the highest degree or level of school you have completed? If currently enrolled, pick the highest degree you have received. \textit{Answer options.} No schooling completed, High school graduate, diploma or the equivalent (for example: GED), Bachelor’s degree, Master’s degree	Doctorate degree, Other, Prefer not to answer
    \item What is your field of study? If you have studied across multiple fields (e.g., undergraduate major/minor or different fields for undergraduate and graduate education) please feel free to select all applicable categories. \textit{Answer options.} Arts and Humanities, e.g., Fine Arts, English Literature, History, Philosophy, etc., Biological Sciences, Agriculture, and Natural Resources, e.g., Biology, Biochemistry, Marine Science, Environmental Studies, etc., Physical Sciences and Mathematics, e.g., Physics, Mathematics, Chemistry, Statistics, etc. (If your field of education is related to computers, please select computer-related fields instead), Computer-related fields, e.g, Computer Science, Information Science, Computer Engineering, etc., Social Sciences, e.g., Sociology, Economics, Psychology, International Relations etc., Business, e.g., Accounting, Business Administration, Management, Marketing, etc., 	Communications, Media, and Public Relations, e.g., Communications, Journalism, Telecommunications, etc., 	Education, e.g., Education, Early Childhood Education, Special Education, etc,	Engineering, e.g., Biomedical Engineering, Electrical Engineering, Civil Engineering, etc. (If your field of education is related to computers, please select computer-related fields instead), Health Professions, e.g., Medicine, Nursing, Pharmacy, Speech Therapy, etc., Social Service Professions, e.g., Military, Forensics, Law, Public Administration, Urban Planning, etc., Other, e.g., Theological Studies, Family Studies, etc., Prefer not to answer
\end{itemize}

\textbf{Demographic Responses} 
Participant demographics are reported in Table \ref{tab:demographics}. \textit{Demographics were statistically similar across conditions (all $ps > .05$). Therefore, demographic variables were not included as covariates in subsequent analyses.}

\begin{table}[hbt!]
\centering
\small
\begin{tabular}{lp{1.2cm}}
\toprule
\textbf{Measure} & \textbf{N} \\
\midrule

\multicolumn{2}{l}{\textbf{Gender}} \\
Male & 252 \\
Female & 246 \\
Non-Binary & 6 \\
Prefer not to answer & 1 \\

\midrule
\multicolumn{2}{l}{\textbf{Age}} \\
18--29 years old & 80 \\
30--39 years old & 147 \\
40--49 years old & 115 \\
50--59 years old & 86 \\
60--69 years old & 53 \\
70--79 years old & 20 \\
80 years or older & 3 \\
Prefer not to answer & 1 \\

\midrule
\multicolumn{2}{l}{\textbf{Native English Speaker}} \\
Yes & 496 \\
No & 9 \\

\midrule
\multicolumn{2}{l}{\textbf{English Proficiency (Non-Native)}} \\
Full bilingual proficiency & 5 \\
Full professional proficiency & 3 \\
Professional working proficiency & 1 \\

\midrule
\multicolumn{2}{l}{\textbf{Race/Ethnicity}} \\
White or Caucasian & 366 \\
Black or African American & 76 \\
Hispanic/Latino & 29 \\
Asian & 23 \\
Other & 7 \\
Prefer not to answer & 4 \\

\midrule
\multicolumn{2}{l}{\textbf{Education}} \\
Bachelor’s degree & 229 \\
High school graduate or equivalent & 143 \\
Master’s degree & 84 \\
Doctorate degree & 22 \\
Other & 22 \\
No schooling completed & 3 \\
Prefer not to answer & 2 \\

\midrule
\multicolumn{2}{l}{\textbf{Field of Education$^{*}$}} \\
Business & 97 \\
Computer-related fields & 60 \\
Arts and Humanities & 41 \\
Health Professions & 38 \\
Social Sciences & 36 \\
Social Service Professions & 22 \\
Biological Sciences/Agriculture & 21 \\
Engineering & 20 \\
Education & 19 \\
Physical Sciences/Mathematics & 13 \\
Communications/Media/Public Relations & 13 \\
Other & 7 \\
Prefer not to answer & 5 \\

\bottomrule
\end{tabular}
\caption{Participant demographics.}
\label{tab:demographics}
\vspace{2mm}

\footnotesize{$^{*}$Participants could select multiple fields of education.}
\end{table}



\subsubsection{Procedures and Measures} \label{appendix: procedures and measures}

\textbf{Onboarding.} After informed consents, participants viewed the following message for onboarding. 

\vspace{0.1in}
\textit{In this study, we aim to understand online user awareness of argumentation that commonly leads to an error in reasoning due to the deceptive nature of its presentation.}

\textit{First, you will be presented with a series of news headline-comment pairs. Then, you will be asked to rate the accuracy of the logical reasoning and indicate your confidence in your answer.}

\textit{You will repeat the procedure for a total of 32 set of texts. After you’re finished, you will be asked to provide your demographic information and answer a few questions regarding your evaluation of the commentator.}

\textit{The entire procedure will take about 30 minutes.}

\textit{NOTE: Throughout this study, once you click the \lq Next\rq \space button, you CANNOT go back to the previous page.}
\vspace{0.1in}

\begin{figure}[hbt!]
\centering
  \fbox{
  \includegraphics[width=0.9\columnwidth]{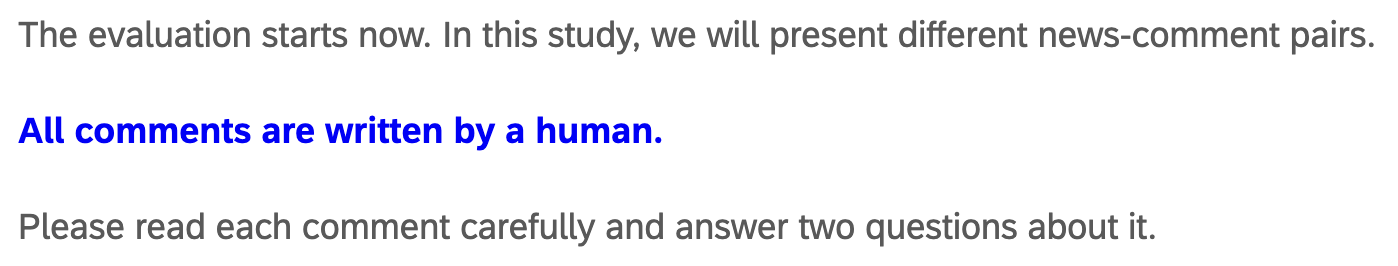}
}
  \caption{Instruction shown to participants before starting the study. The sentence highlighted in \textcolor{blue}{blue} was presented differently according to the groups the participants were assigned to.}
  \label{fig:instruction}
\end{figure}

Next, participants were informed about how to receive payment from Prolific and that they would encounter two attention checks in the study. Failure to answer the attention check questions resulted in termination with a payment of 0.2\$, although Prolific permitted no payment in these cases. They were also informed about duplicate submissions, as duplicate submissions were prohibited in this study. Next, all participants were randomly assigned to a source condition (control, Human, Human+AI, AI+Human, AI) and received instructions based on which group they were assigned to (Figure \ref{fig:instruction}).

\begin{figure}[hbt!]
\centering
\includegraphics[width=\columnwidth]{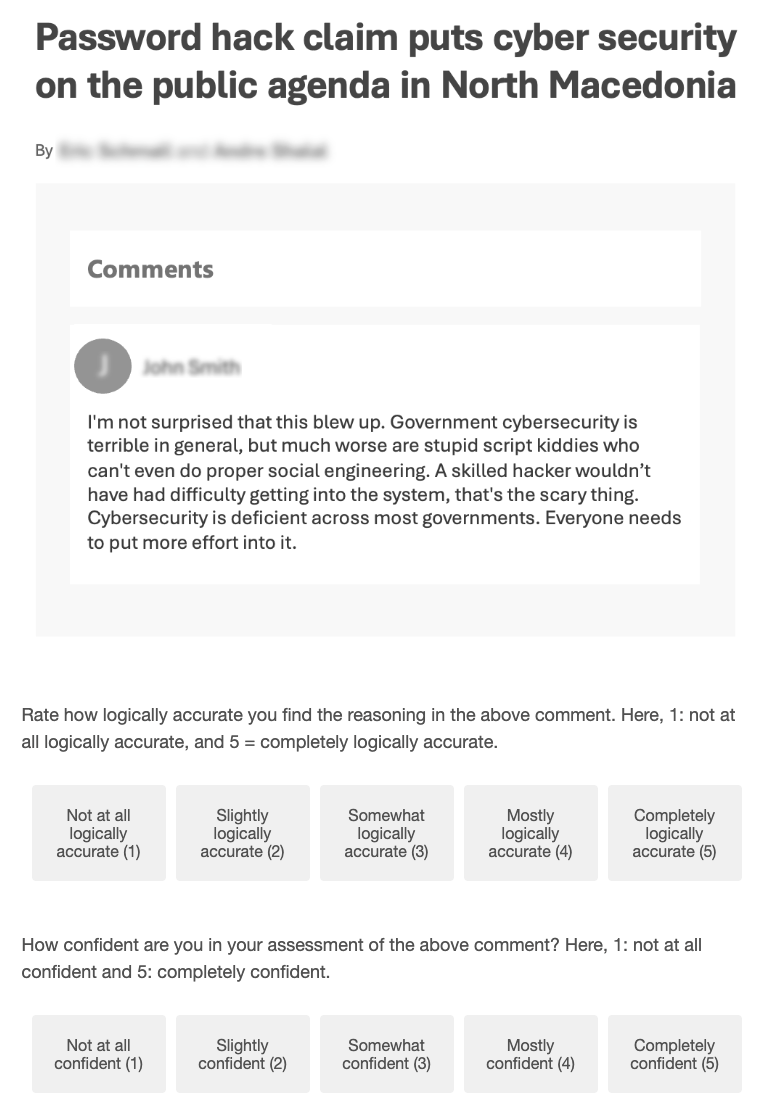}
  \caption{An example of a stimulus shown to participants. For each stimulus, participants were asked to rate logical accuracy and their confidence in their logical accuracy judgments in 5-point scales.}
  \label{fig:logical_accuracy_confidence}
\end{figure}

\textbf{Main study.} During the main study, each participant evaluated 32 news headline-comment pairs and answered questions about the logical accuracy of the comments and their confidence in their logical accuracy ratings (Figure \ref{fig:logical_accuracy_confidence}). 

\begin{figure}[hbt!]
\centering
\includegraphics[width=\columnwidth]{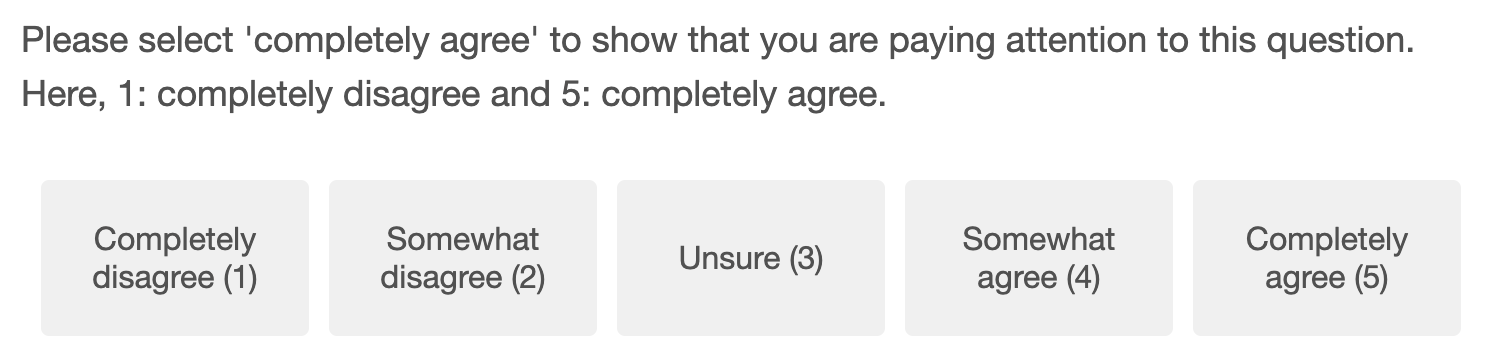}
  \caption{Attention-check question.}
  \label{fig:attention_check}
\end{figure}

During the evaluation, participants were exposed to two attention-check questions, as shown in Figure \ref{fig:attention_check}. The study automatically terminated for participants who failed either of the attention-check questions. 

\vspace{0.1in}
\textbf{Manipulation check question.} After completing the main study, participants answered a manipulation check question. \textit{Please indicate which source condition you encountered in this study.} Answer options: Source wasn't mentioned directly, Written by human, Written by AI, Written by human with AI assistance, Written by AI with human assistance. The manipulation check was used to ensure that participants noticed and correctly interpreted the source labels presented during the study \cite{chen2023ai}.

\vspace{0.1in}
\textbf{Demographic questions.} The demographic questions and participants' responses have been reported in Appendix \ref{appendix: participant demographic}.

\vspace{0.1in}
\textbf{Computer and AI expertise questions.} After answering demographic questions, participants answered some computer and AI expertise questions. This allowed us to examine whether there were any meaningful differences across source conditions.

\begin{itemize}
    \item  How would you rate your computer expertise on a 5-point scale? Here, 1 means novice and 5 means expert. \textit{Answer options.} Novice (1), Basic (2), Intermediate (3), Advanced (4), Expert (5)
    \item How would you rate your level of expertise with artificial intelligence (AI) tools (e.g., ChatGPT, Gemini, Copilot, Midjourney, etc.) on a 5-point scale? Here, 1 means no experience and 5 means expert. \textit{Answer options.} No experience (1), Basic understanding, i.e., familiar with common AI tools or concepts (2), Intermediate, i.e., can use AI tools or understand key ideas with some confidence (3), Advanced, i.e., can apply AI methods or tools effectively in work or study (4), Expert, i.e., possess deep knowledge or professional experience with AI systems or research (5)
    \item How frequently do you use artificial intelligence (AI) tools (e.g., ChatGPT, Gemini, Copilot, Midjourney, etc.) on a 5-point scale? \textit{Answer options.} Never (1), Rarely, i.e., a few times a year (2), Occasionally, i.e., a few times a month (3), Frequently, i.e., a few times a week (4), Very frequently, i.e., daily or almost daily (5)

\end{itemize}

\textbf{News source preference and political alignment question.} Participants were asked a question that allowed us to determine their preferences in terms of news source. While this study mainly focuses on online news media, it is important to understand participants' preferences in order to better understand the biases that occurred in the results. \textit{Which of the following best describes your most preferred source of news? Please feel free to select multiple options if applicable.} Answer options. Television news channels; News websites and apps; Social media platforms such as Instagram, Facebook, X, etc.; Search engines such as Google, Bing, etc.; Podcasts or radio; Newspapers	Government websites	Friends, family, or word of mouth; Others, please specify (Text box).

We also asked participants to answer a question about their political alignment. While we utilized a two-step verification process (manual and LLM-based) to ensure that the news headlines and associated comments were politically and emotionally neutral in the context of the United States, it is still important to understand if there are any differences in participants' political alignments across conditions to better situate the associated biases. \textit{How will you describe your political alignment? Here, 1 means extremely left and 5 means extremely right.
These answers will not be traced back to you or impact your compensation, so please answer honestly.} Answer options. Completely left (1), Somewhat left (2), Neutral; neither left nor right (3), Somewhat right (4), Completely right (5).

\vspace{0.1in}
\textbf{Trust and evaluation questions.} We also asked participants to rate their trust in \cite{chen2023ai, zhou2025effect} and evaluation of the commenters \cite{nahar2025catch}, as source labels may impact downstream processes such as trust and evaluation. Trust was measured across cognitive, affective, and behavioral dimensions \cite{johnson2005cognitive, schlenker1973effects}. Participants were asked: 

\vspace{0.1in}
\textit{Please indicate how well each sentence describes your opinion about the commenter whose comments you evaluated: I am confident in this commentator (C); This commentator has integrity (C); This commentator is reliable (C); I can trust this commentator (C); I am familiar with this commentator’s style/content (C); I feel this commentator communicates in a caring or thoughtful manner (A); This commentator displays a warm and caring attitude (A); I am willing to rely on this commentator when forming my own opinions (B); I am willing to consider information conveyed by this commentator when forming my own opinions (B). Note. C = cognitive trust; A = affective trust; B = behavioral trust). For each measure, participants chose from one of the following options. Doesn’t describe at all (1), Doesn’t describe much (2), Describes somewhat (3)	Describes mostly (4), Describes very well (5)}.

\vspace{0.1in}
Evaluation measures focused on warmth and competence \cite{cuddy2008warmth, fiske2018model}. Participants were asked: 

\vspace{0.1in}
\textit{Please indicate how well each adjective describes the commenter you’ve evaluated comments from: Likable, Friendly, Pleasant, Competent, Intelligent, Capable, Efficient, Helpful.} For each measure, participants chose from one of the following options. \textit{Doesn’t describe at all (1), Doesn’t describe much (2), Describes somewhat (3)	Describes mostly (4), Describes very well (5)}.
\begin{figure}[hbt!]
\centering
\includegraphics[width=\columnwidth]{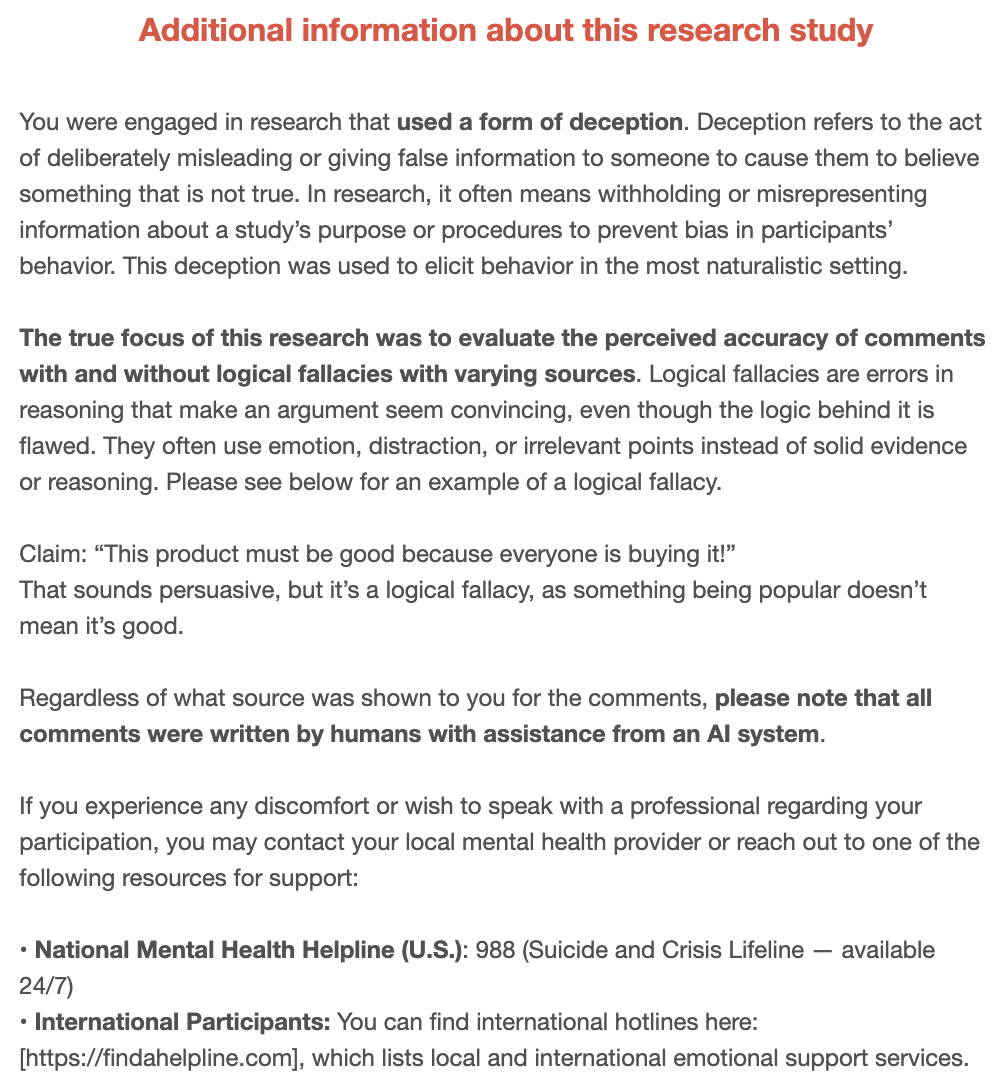}
  \caption{Debrief information shown to participants at the end of the study.}
  \label{fig:debrief}
\end{figure}

\textbf{Process of comment writing question.} To understand participants' perceptions regarding how the comments were written, we asked them: \textit{How do you think the comments you evaluated were written?}

\vspace{0.1in}
\textbf{Source preference rank questions.} Participants were also asked to rank the sources present in the study, in order to understand whether participants' behavioral patterns followed their explicit preferences. 

\vspace{0.1in}\textit{Please rank the following comment sources according to your preference.} Rank options. Human; Human with AI assistance; AI; AI with Human assistance; No source disclosure. 

\vspace{0.1in}Participants were also asked, 

\vspace{0.1in}\textit{Why did you rank the comment sources in that order?} to better understand why they ranked the sources in the ways that they did.

\vspace{0.1in}
\textbf{Debrief information.} As the study contained elements of deception with manipulated source disclosure, we included a debrief document at the end of the study, in accordance with IRB guidelines (Figure \ref{fig:debrief}).


\subsection{LLM Evaluation} \label{appendix: method_llm_evaluation}
To enable a direct comparison with human judgments, we evaluate three state-of-the-art LLMs: GPT-5.2 \citep{singh2025gpt5}, Gemini 2.5 Flash \cite{comanici2025gemini2.5}, and Claude Sonnet 4.5 \citep{anthropic2025claudesonnet45} on the same set of comments. The prompt format and rating scales are identical to those presented to humans. 

Models were presented with identical stimuli and instructed to rate perceived logical accuracy and confidence on a 5-point scale, consistent with the human study. We used the models through their official API with identical settings (temperature=0) for each model at the time of evaluation to make the results reproducible, and each stimulus was evaluated independently to avoid context carryover across inputs. Also, we perform the evaluation three times separately for each model to reduce random error and assess the consistency of the LLM-as-judge evaluation \citep{gu2026survey, pan2024human}. This setup allows direct comparison between human and model judgments under identical conditions. As trust in the commenter and overall evaluation capture interpersonal social judgments that are most meaningful for humans, these measures were only collected in the human study. The prompt is as follows.

\vspace{0.1in}
\textit{You will be presented with a news headline-comment pair. Then, you will be asked to rate the accuracy of the logical reasoning and indicate your confidence in your answer.}

\textit{[If source disclosed] All comments are written by a [human/AI/human with AI assistance/AI with human assistance].}

\textit{Read the comment carefully and answer two questions about it. 
}

\textit{Rate how logically accurate you find the reasoning in the comment. Here, 1 = not at all logically accurate, 2= slightly logically accurate, 3 = somewhat logically accurate, 4 = mostly logically accurate, and 5 = completely logically accurate.}

\textit{How confident are you in your assessment of the comment? Here, 1 = not at all confident, 2 = slightly confident, 3 = somewhat confident, 4 = mostly confident, and 5 = completely confident.}


\subsection{Data Analysis} \label{appendix: data analysis}
We used linear mixed-effects regression (LMER) models to analyze the data. The LMER models included participant, item, and stimulus set as random intercepts to account for repeated observations and stimulus-level variability. For perceived logical accuracy, we used the model: lmer(Accuracy ~ Condition * Fallacy + (1 | Subject) + (1 | ItemID) + (1 | SetID), data = data). For confidence, we used the model: lmer(Confidence ~ Condition * Fallacy + (1 | Subject)  + (1 | ItemID) + (1 | SetID), data = data). For trust: lmer(Trust ~ Condition + (1 | Subject) + (1 | SetID), data = data). Finally, for overall evaluation: lmer(Evaluation ~ Condition + (1 | Subject) + (1 | SetID), data = data).

\subsection{Results: Perceived Logical Accuracy} \label{appendix: results_perceived_logical_accuracy}

\begin{table*}[t]
\centering
\small
\begin{tabular}{llccccc}
\toprule
\textbf{Group} & \textbf{Fallacy} & \textbf{Control} & \textbf{Human} & \textbf{AI} & \textbf{Human+AI} & \textbf{AI+Human} \\
\midrule

\multirow{2}{*}{Human Evaluators}
& Absent 
& 3.87 [.134] 
& 3.77 [.139] 
& 3.81 [.130] 
& 3.86 [.132] 
& 3.88 [.129] \\

& Present 
& 2.40 [.136] 
& 3.35 [.141] 
& 2.53 [.129] 
& 3.37 [.138] 
& 2.59 [.134] \\

\midrule

\multirow{2}{*}{GPT-5.2}
& Absent 
& 3.52 [.136] 
& 3.50 [.136] 
& 3.23 [.136] 
& 3.38 [.136] 
& 3.27 [.136] \\

& Present 
& 2.72 [.136] 
& 2.83 [.136] 
& 2.58 [.136] 
& 2.66 [.136] 
& 2.62 [.136] \\

\midrule

\multirow{2}{*}{Gemini 2.5 Flash}
& Absent 
& 3.67 [.132] 
& 3.67 [.132] 
& 3.46 [.132] 
& 3.65 [.132] 
& 3.52 [.132] \\

& Present 
& 2.33 [.132] 
& 2.41 [.132] 
& 2.29 [.132] 
& 2.43 [.132] 
& 2.29 [.132] \\

\midrule

\multirow{2}{*}{Claude Sonnet 4.5}
& Absent 
& 2.94 [.140] 
& 3.00 [.140] 
& 2.67 [.140] 
& 2.88 [.140] 
& 2.75 [.140] \\

& Present 
& 2.20 [.140] 
& 2.22 [.140] 
& 2.18 [.140] 
& 2.20 [.140] 
& 2.16 [.140] \\

\bottomrule
\end{tabular}
\caption{Estimated marginal means for perceived logical accuracy across source conditions and fallacy presence. Note. [] indicates standard error.}
\label{tab:logical_accuracy_means}
\end{table*}

\subsubsection{Descriptive Statistics} \label{appendix: pa_descriptive_stat}
Table~\ref{tab:logical_accuracy_means} presents the estimated marginal means for perceived logical accuracy across source conditions and fallacy presence. For human evaluators, ratings for non-fallacious comments were relatively similar across conditions, whereas substantial differences emerged for fallacious comments. In particular, fallacious comments received notably higher logical accuracy ratings in the Human ($M=3.35$) and Human+AI ($M=3.37$) conditions compared to the Control ($M=2.40$), AI ($M=2.53$), and AI+Human ($M=2.59$) conditions, consistent with the observed Condition $\times$ Fallacy interaction. In contrast, LLMs showed comparatively stable ratings across source conditions, while still assigning lower ratings to fallacious comments overall. Among the evaluated models, Gemini 2.5 Flash demonstrated the largest separation between fallacious and non-fallacious comments, suggesting greater sensitivity to reasoning quality.

\subsubsection{Fallacy Penalty} \label{appendix: pa_fallacy_penalty}

\begin{table}[t]
\centering
\small
\begin{tabular}{lc}
\toprule
\textbf{Group / Condition} & \textbf{Fallacy Penalty} \\
\midrule

Human Evaluators -- Control & 1.47 \\
Human Evaluators -- Human & \textbf{0.42} \\
Human Evaluators -- AI & 1.28 \\
Human Evaluators -- Human+AI & \textbf{0.49} \\
Human Evaluators -- AI+Human & 1.29 \\

\midrule

GPT-5.2 -- Control & 0.80 \\
GPT-5.2 -- Human & 0.67 \\
GPT-5.2 -- AI & 0.65 \\
GPT-5.2 -- Human+AI & 0.72 \\
GPT-5.2 -- AI+Human & 0.65 \\

\midrule

Gemini 2.5 Flash -- Control & 1.34 \\
Gemini 2.5 Flash -- Human & 1.26 \\
Gemini 2.5 Flash -- AI & 1.17 \\
Gemini 2.5 Flash -- Human+AI & 1.22 \\
Gemini 2.5 Flash -- AI+Human & 1.23 \\

\midrule

Claude Sonnet 4.5 -- Control & 0.74 \\
Claude Sonnet 4.5 -- Human & 0.78 \\
Claude Sonnet 4.5 -- AI & 0.49 \\
Claude Sonnet 4.5 -- Human+AI & 0.68 \\
Claude Sonnet 4.5 -- AI+Human & 0.59 \\

\bottomrule
\end{tabular}
\caption{Fallacy penalty across groups and source conditions, calculated as the difference between non-fallacy and fallacy perceived logical accuracy ratings. A larger penalty indicates that fallacious comments were rated much lower than non-fallacious ones, i.e., the evaluator was more sensitive to logical flaws. A smaller penalty indicates that fallacious comments were rated closer to non-fallacious comments, i.e., the evaluator was more susceptible to flawed reasoning.}
\label{tab:fallacy_penalty}
\end{table}

The fallacy penalty was calculated as the difference between perceived logical accuracy ratings for non-fallacious and fallacious comments. Larger fallacy penalties therefore indicate stronger penalization of logically flawed comments and greater sensitivity to reasoning errors. Table~\ref{tab:fallacy_penalty} shows that human evaluators exhibited substantially smaller fallacy penalties in the Human (0.42) and Human+AI (0.49) conditions compared to the Control (1.47), AI (1.28), and AI+Human (1.29) conditions, suggesting greater susceptibility to flawed reasoning when comments were associated with human involvement. In contrast, LLMs showed relatively stable fallacy penalties across source conditions, consistent with the absence of significant source-label interactions. Although Claude Sonnet 4.5 had a smaller fallacy penalty (0.49) in the AI condition, the differences with other conditions were not statistically significant after Tukey adjustment. Among the evaluated models, Gemini 2.5 Flash exhibited the largest fallacy penalties overall, indicating the greatest sensitivity to logical flaws, whereas Claude Sonnet 4.5 showed comparatively smaller penalties.

\subsubsection{Pairwise Comparisons} \label{appendix_pa_pairwise}
\begin{table*}[t]
\centering
\small
\begin{tabular}{llccc}
\toprule
\textbf{Group} & \textbf{Comparison} & \textbf{Mean Diff.} & \textbf{$t$} & \textbf{$p_{adj}$} \\
\midrule

\multicolumn{5}{l}{\textbf{LLMs -- Non-Fallacies}} \\
\midrule
& No significant comparisons & -- & -- & -- \\
\midrule

\multicolumn{5}{l}{\textbf{Human Evaluators -- Fallacies}}  \\
\midrule
& \textbf{Control vs Human} & \textbf{$-0.955$} & \textbf{$-11.28$} & \textbf{$<.001$} \\
& Control vs AI & $-0.133$ & $-1.57$ & $.518$ \\
& \textbf{Control vs Human+AI} & \textbf{$-0.977$} & \textbf{$-11.46$} & \textbf{$<.001$} \\
& Control vs AI+Human & $-0.194$ & $-2.28$ & $.154$ \\
& \textbf{Human vs AI} & \textbf{$0.822$} & \textbf{$9.83$} & \textbf{$<.001$} \\
& Human vs Human+AI & $-0.022$ & $-0.27$ & $.999$ \\
& \textbf{Human vs AI+Human} & \textbf{$0.761$} & \textbf{$9.06$} & \textbf{$<.001$} \\
& \textbf{AI vs Human+AI} & \textbf{$-0.844$} & \textbf{$-10.02$} & \textbf{$<.001$} \\
& AI vs AI+Human & $-0.061$ & $-0.72$ & $.951$ \\
& \textbf{Human+AI vs AI+Human} & \textbf{$0.783$} & \textbf{$9.26$} & \textbf{$<.001$} \\

\midrule

\multicolumn{5}{l}{\textbf{LLMs -- Non-Fallacies}} \\
\midrule
& Claude Sonnet 4.5: No significant comparisons & -- & -- & -- \\
& Gemini 2.5 Flash: No significant comparisons & -- & -- & -- \\
& GPT-5.2: No significant comparisons & -- & -- & -- \\

\midrule

\multicolumn{5}{l}{\textbf{LLMs -- Fallacies}} \\
\midrule
& Claude Sonnet 4.5: No significant comparisons & -- & -- & -- \\
& Gemini 2.5 Flash: No significant comparisons & -- & -- & -- \\
& GPT-5.2: No significant comparisons & -- & -- & -- \\

\bottomrule
\end{tabular}
\caption{Pairwise comparisons for perceived logical accuracy across source conditions within human and LLM groups (Tukey-adjusted). Significant comparisons are shown in bold.}
\label{tab:appendix_logical_pairwise}
\end{table*}

Table~\ref{tab:appendix_logical_pairwise} presents the Tukey-adjusted pairwise comparisons for perceived logical accuracy across source conditions. For non-fallacious comments, human evaluators showed no significant differences across source conditions, indicating that source labels had little influence when reasoning quality was high. In contrast, substantial differences emerged for fallacious comments. Specifically, the Human and Human+AI conditions received significantly higher logical accuracy ratings than the Control, AI, and AI+Human conditions (all $p_{adj}<.001$), consistent with the observed Condition $\times$ Fallacy interaction. In contrast, LLMs showed no significant pairwise differences across source conditions for either fallacious or non-fallacious comments, further supporting the finding that LLM judgments were comparatively stable across source-label manipulations.

\begin{table*}[t]
\centering
\small
\begin{tabular}{llccc}
\toprule
\textbf{Group} & \textbf{Condition} & \textbf{Mean Diff.} & \textbf{$t$} & \textbf{$p_{adj}$} \\
\midrule

\multicolumn{5}{l}{\textbf{Human Evaluators}} \\
\midrule
& Control & 1.47 & 28.85 & $<.001$ \\
& Human & 0.42 & 8.44 & $<.001$ \\
& AI & 1.28 & 25.68 & $<.001$ \\
& Human+AI & 0.49 & 9.71 & $<.001$ \\
& AI+Human & 1.29 & 25.76 & $<.001$ \\

\midrule

\multicolumn{5}{l}{\textbf{Claude Sonnet 4.5}} \\
\midrule
& Control & 0.735 & 4.25 & $<.001$ \\
& Human & 0.776 & 4.49 & $<.001$ \\
& AI & 0.485 & 2.81 & $.005$ \\
& Human+AI & 0.672 & 3.89 & $<.001$ \\
& AI+Human & 0.589 & 3.41 & $<.001$ \\

\midrule

\multicolumn{5}{l}{\textbf{Gemini 2.5 Flash}} \\
\midrule
& Control & 1.339 & 7.75 & $<.001$ \\
& Human & 1.256 & 7.27 & $<.001$ \\
& AI & 1.172 & 6.79 & $<.001$ \\
& Human+AI & 1.214 & 7.03 & $<.001$ \\
& AI+Human & 1.235 & 7.15 & $<.001$ \\

\midrule

\multicolumn{5}{l}{\textbf{GPT-5.2}} \\
\midrule
& Control & 0.797 & 4.62 & $<.001$ \\
& Human & 0.672 & 3.89 & $<.001$ \\
& AI & 0.651 & 3.77 & $<.001$ \\
& Human+AI & 0.714 & 4.13 & $<.001$ \\
& AI+Human & 0.651 & 3.77 & $<.001$ \\

\bottomrule
\end{tabular}
\caption{Pairwise comparisons between non-fallacious and fallacious comments within each evaluator group and source condition (Tukey-adjusted). Positive values indicate higher perceived logical accuracy ratings for non-fallacious comments.}
\label{tab:appendix_fallacy_vs_nonfallacy}
\end{table*}

Table~\ref{tab:appendix_fallacy_vs_nonfallacy} presents Tukey-adjusted pairwise comparisons between non-fallacious and fallacious comments within each evaluator group and source condition. Human evaluators consistently rated non-fallacious comments as more logically accurate than fallacious comments across all source conditions ($p<.001$). Similarly, all LLMs assigned significantly lower ratings to fallacious comments across every source condition. Among the evaluated models, Gemini 2.5 Flash showed the largest differences between fallacious and non-fallacious comments, suggesting greater sensitivity to logical flaws, whereas Claude Sonnet 4.5 exhibited comparatively smaller differences.

\begin{table}[t]
\centering
\small
\begin{tabular}{llc}
\toprule
\textbf{Comparison} & \textbf{$t$} & \textbf{$p$} \\
\midrule

\multicolumn{3}{l}{\textbf{Fallacious Comments}} \\
\midrule
Humans vs GPT-5.2 & 6.70 & $<.001$ \\
Humans vs Claude & 15.94 &  $<.001$ \\
Humans vs Gemini & 9.07 &  $<.001$ \\
GPT-5.2 vs Claude & 6.53 &  $<.001$ \\
GPT-5.2 vs Gemini & 3.44 &  $.001$ \\
Claude vs Gemini & $-1.61$ &  $.109$ \\

\midrule

\multicolumn{3}{l}{\textbf{Non-Fallacious Comments}} \\
\midrule
Humans vs GPT-5.2 & 4.01 &  $<.001$ \\
Humans vs Claude & 12.05 &  $<.001$ \\
Humans vs Gemini & $-0.23$ &  $.820$ \\
GPT-5.2 vs Claude & 7.79 &  $<.001$ \\
GPT-5.2 vs Gemini & $-2.24$ &  $.052$ \\
Claude vs Gemini & $-7.35$ &  $<.001$ \\

\bottomrule
\end{tabular}
\caption{Pairwise comparisons between human and LLMs for perceived logical accuracy ratings.}
\label{tab:human_llm_pairwise}
\end{table}

Pairwise comparisons of logical accuracy ratings between human and LLMs are presented in Table~\ref{tab:human_llm_pairwise}. Welch two-sample $t$-tests with Holm correction were used due to unequal variances across humans and LLMs. For fallacious comments, humans assigned significantly higher logical accuracy ratings than all evaluated LLMs (all $ps<.001$). Among LLMs, GPT-5.2 assigned significantly higher ratings to fallacious comments than both Gemini (\(p=.001\)) and Claude (\(p<.001\)), while Gemini and Claude did not differ significantly (\(p=.109\)). For non-fallacious comments, humans rated comments similarly to Gemini, but significantly higher than GPT-5.2 and Claude. Claude generally assigned the lowest ratings across evaluators, whereas Gemini produced ratings most similar to humans for non-fallacious content.

\subsubsection{Fallacy-Type Analysis} \label{appendix:pa_fallacy_type_analysis}
\begin{table*}[t]
\centering
\small
\begin{tabular}{p{1.4cm}p{7cm}c}
\toprule
\textbf{Evaluator} & \textbf{Comparison} & \textbf{$p$} \\
\midrule

\multirow{12}{*}{Human}
& Hasty generalization $>$ Appeal to tradition & $<.001$ \\
& Hasty generalization $>$ Appeal to worse problems & $<.001$ \\
& Hasty generalization $>$ Appeal to majority & $<.001$ \\
& Hasty generalization $>$ Slippery slope & $<.001$ \\
& False dilemma $>$ Appeal to authority & $<.001$ \\
& False dilemma $>$ Appeal to majority & $<.001$ \\
& False dilemma $>$ Slippery slope & $<.001$ \\
& Appeal to authority $<$ Appeal to nature & $<.001$ \\
& Appeal to majority $<$ Appeal to nature & $<.001$ \\
& Appeal to majority $<$ Appeal to tradition & $<.001$ \\
& Slippery slope $<$ Appeal to tradition & $<.001$ \\
& Slippery slope $<$ Appeal to nature & $<.001$ \\

\midrule

\multirow{12}{*}{LLMs}
& Appeal to nature $>$ Appeal to authority & $<.001$ \\
& Appeal to nature $>$ Appeal to majority & $<.001$ \\
& Appeal to tradition $>$ Appeal to authority & $<.001$ \\
& Appeal to tradition $>$ Appeal to majority & $<.001$ \\
& Appeal to worse problems $<$ Appeal to nature & $<.001$ \\
& Appeal to worse problems $<$ Appeal to tradition & $<.001$ \\
& False dilemma $<$ Appeal to nature & $<.001$ \\
& False dilemma $<$ Appeal to tradition & $<.001$ \\
& Hasty generalization $<$ Appeal to nature & $<.001$ \\
& Hasty generalization $<$ Appeal to tradition & $<.001$ \\
& Slippery slope $<$ Appeal to nature & $<.001$ \\
& Slippery slope $<$ Appeal to tradition & $<.001$ \\

\bottomrule
\end{tabular}
\caption{Pairwise comparisons across logical fallacy types for perceived logical accuracy ratings. Only selected statistically significant comparisons are shown for readability.}
\label{tab:appendix_pa_fallacy_type_pairwise}
\end{table*}

Both humans and LLMs varied in perceived logical accuracy across fallacy types (Table \ref{tab:appendix_pa_fallacy_type_pairwise}; Welch two-sample $t$-tests with Holm correction were used due to unequal variances across groups). This suggests that some reasoning errors were more difficult to detect than others. Human evaluators assigned relatively higher logical accuracy ratings to \textit{hasty generalization} and \textit{false dilemma} compared to several other fallacy types, while \textit{slippery slope} and \textit{appeal to majority} generally received lower ratings ($p<.001$). In contrast, LLMs assigned the highest logical accuracy ratings to \textit{appeal to nature} and \textit{appeal to tradition} ($p<.001$), whereas \textit{appeal to worse problems}, \textit{false dilemma}, and \textit{slippery slope} tended to receive lower ratings. Interestingly, the differing error patterns between humans and LLMs suggest complementary strengths, highlighting the potential value of human--LLM collaboration for reasoning evaluation tasks.

\subsection{Results: Confidence in Logical Accuracy}

\subsubsection{Descriptive Statistics} \label{appendix: confidence_descriptive_statistics}

\begin{table*}[t]
\centering
\small
\begin{tabular}{llccccc}
\toprule
\textbf{Group} & \textbf{Fallacy} & \textbf{Control} & \textbf{Human} & \textbf{AI} & \textbf{Human+AI} & \textbf{AI+Human} \\
\midrule

\multirow{2}{*}{Human Evaluators}
& Absent 
& 3.93 [.071] 
& 3.78 [.062] 
& 3.92 [.063] 
& 3.92 [.074] 
& 3.95 [.081] \\

& Present 
& 3.82 [.063] 
& 3.79 [.084] 
& 3.87 [.081] 
& 3.89 [.062] 
& 3.90 [.069] \\

\midrule

\multirow{2}{*}{GPT-5.2}
& Absent 
& 4.34 [.063] 
& 4.42 [.063] 
& 4.42 [.063] 
& 4.40  [.063]  
& 4.48  [.063]  \\

& Present 
& 4.29 [.063] 
& 4.29  [.063] 
& 4.46  [.063] 
& 4.38  [.063] 
& 4.46  [.063]  \\

\midrule

\multirow{2}{*}{Gemini 2.5 Flash}
& Absent 
& 4.71 [.058] 
& 4.65 [.058] 
& 4.71 [.058] 
& 4.50 [.058] 
& 4.57 [.058]  \\

& Present 
& 4.67 [.058] 
& 4.60 [.058] 
& 4.67 [.058] 
& 4.67 [.058] 
& 4.67 [.058]  \\

\midrule

\multirow{2}{*}{Claude Sonnet 4.5}
& Absent 
& 4.04 [.067] 
& 4.02 [.067] 
& 4.07 [.067] 
& 4.04 [.067]  
& 4.07 [.067]  \\

& Present 
& 4.27 [.067] 
& 4.25 [.067] 
& 4.29 [.067] 
& 4.29 [.067] 
& 4.31 [.067]  \\

\bottomrule
\end{tabular}
\caption{Estimated marginal means for confidence in logical accuracy across source conditions and fallacy presence. Note. [] indicates standard error.}
\label{tab:confidence_means}
\end{table*}

Table~\ref{tab:confidence_means} presents the estimated marginal means for confidence in logical accuracy judgments across source conditions and fallacy presence. Human evaluators showed relatively stable confidence ratings across both source conditions and fallacy presence. LLMs were also consistent across source conditions, but they consistently assigned higher confidence scores than humans across all conditions, with Gemini 2.5 Flash exhibiting the highest confidence overall, followed by GPT-5.2 and Claude Sonnet 4.5. Interestingly, whereas GPT-5.2 and Gemini showed only minor confidence differences between fallacious and non-fallacious comments, Claude Sonnet 4.5 exhibited slightly higher confidence for fallacious comments. Overall, the results suggest that LLM confidence judgments were comparatively stable across source conditions, but varied substantially across models, with all evaluated LLMs demonstrating consistently high confidence even when evaluating logically fallacious content.

\subsubsection{Pairwise Comparisons} \label{appendix: confidence_pairwise_comparisons}

For human confidence ratings, most pairwise comparisons were not significant after Tukey correction, other than fallacy vs. no-fallacy in the control condition, where participants were less confident when evaluating fallacious than non-fallacious comments ($p<.001$). LLMs also exhibited mostly stable confidence ratings, other than two small pairwise differences for Gemini 2.5 Flash in the non-fallacy condition, where the Control and AI conditions showed slightly higher confidence ratings than the Human+AI condition ($p_{adj}=.025$); no other pairwise comparisons were significant. However, these effects were comparatively small and did not form a consistent pattern across models or source conditions. 

\begin{table*}[t]
\centering
\small
\begin{tabular}{p{1.8cm}p{3.0cm}ccc}
\toprule
\textbf{Group} & \textbf{Condition} & \textbf{Diff.} & \textbf{$t$} & \textbf{$p_{adj}$} \\
\midrule

\multirow{5}{*}{Humans}
& Control & 0.112 & 3.51 & $<.001$ \\
& Human & $-0.013$ & $-0.43$ & $.667$ \\
& AI & 0.051 & 1.64 & $.101$ \\
& Human+AI & 0.030 & 0.95 & $.341$ \\
& AI+Human & 0.047 & 1.50 & $.134$ \\

\midrule

\multirow{5}{*}{Claude}
& Control & $-0.227$ & $-2.59$ & $.010$ \\
& Human & $-0.227$ & $-2.59$ & $.010$ \\
& AI & $-0.227$ & $-2.59$ & $.010$ \\
& Human+AI & $-0.248$ & $-2.83$ & $.005$ \\
& AI+Human & $-0.248$ & $-2.83$ & $.005$ \\

\midrule

\multirow{5}{*}{Gemini}
& Control & 0.044 & 0.50 & $.619$ \\
& Human & 0.044 & 0.50 & $.619$ \\
& AI & 0.044 & 0.50 & $.619$ \\
& Human+AI & $-0.165$ & $-1.88$ & $.061$ \\
& AI+Human & $-0.102$ & $-1.16$ & $.245$ \\

\midrule

\multirow{5}{*}{GPT-5.2}
& Control & 0.044 & 0.50 & $.619$ \\
& Human & 0.127 & 1.45 & $.148$ \\
& AI & $-0.040$ & $-0.45$ & $.651$ \\
& Human+AI & 0.023 & 0.26 & $.795$ \\
& AI+Human & 0.023 & 0.26 & $.795$ \\

\bottomrule
\end{tabular}
\caption{Pairwise comparisons between non-fallacious and fallacious comments for confidence ratings across evaluator groups and source conditions (Tukey-adjusted). Positive values indicate higher confidence for non-fallacious comments.}
\label{tab:confidence_fallacy_diff}
\end{table*}

Table~\ref{tab:confidence_fallacy_diff} presents pairwise comparisons between confidence ratings for non-fallacious and fallacious comments across evaluator groups and source conditions. Human evaluators showed only a small confidence difference between non-fallacious and fallacious comments, with a significant effect emerging only in the Control condition. Similarly, GPT-5.2 and Gemini 2.5 Flash exhibited largely stable confidence ratings across fallacy presence. In contrast, Claude Sonnet 4.5 consistently showed slightly higher confidence for fallacious comments across conditions. Overall, for human evaluators, confidence judgments appeared substantially less sensitive to fallacy presence than perceived logical accuracy ratings.

\begin{table}[t]
\centering
\small
\begin{tabular}{llc}
\toprule
\textbf{Comparison} & \textbf{$t$} & \textbf{$p$} \\
\midrule

\multicolumn{3}{l}{\textbf{Fallacious Comments}} \\
\midrule
Humans vs GPT-5.2 & $-12.33$ & $<.001$ \\
Humans vs Claude & $-10.34$ & $<.001$ \\
Humans vs Gemini & $-18.90$ & $<.001$ \\

\midrule

\multicolumn{3}{l}{\textbf{Non-Fallacious Comments}} \\
\midrule
Humans vs GPT-5.2 & $-12.15$ & $<.001$ \\
Humans vs Claude & $-4.82$ & $<.001$ \\
Humans vs Gemini & $-17.32$ & $<.001$ \\

\bottomrule
\end{tabular}
\caption{Pairwise comparisons between human and LLMs for confidence ratings.}
\label{tab:human_llm_confidence_pairwise}
\end{table}

Table~\ref{tab:human_llm_confidence_pairwise} presents pairwise comparisons between human and LLMs for confidence ratings. Welch two-sample $t$-tests with Holm correction were used due to unequal variances across groups. Across both fallacious and non-fallacious comments, all evaluated LLMs assigned significantly higher confidence scores than human evaluators ($p<.001$). Gemini 2.5 Flash exhibited the highest confidence overall, followed by GPT-5.2 and Claude Sonnet 4.5. These findings suggest that LLMs maintained consistently high confidence across reasoning tasks, even when evaluating logically fallacious content.

\subsubsection{Fallacy-Type Analysis} \label{appendix: confidence_fallacy_type_analysis}
\begin{table*}[t]
\centering
\small
\begin{tabular}{p{1.6cm}p{7.2cm}c}
\toprule
\textbf{Evaluator} & \textbf{Comparison} & \textbf{$p$} \\
\midrule

\multirow{12}{*}{Human}
& Slippery slope $>$ False dilemma & $<.01$ \\
& Slippery slope $>$ Appeal to tradition & $<.01$ \\
& Slippery slope $>$ Appeal to nature & $<.01$ \\
& Appeal to worse problems $>$ Hasty generalization & $<.01$ \\
& Appeal to worse problems $>$ False dilemma & $<.01$ \\
& Appeal to majority $>$ False dilemma & $<.01$ \\
& Appeal to majority $>$ Appeal to tradition & $<.01$ \\
& Appeal to majority $>$ Appeal to nature & $<.01$ \\
& Appeal to authority $>$ Appeal to nature & $<.01$ \\
& Hasty generalization $<$ Slippery slope & $<.01$ \\
& Hasty generalization $<$ Appeal to tradition & $<.01$ \\
& Hasty generalization $>$ Appeal to majority & $<.01$ \\

\midrule

\multirow{12}{*}{LLMs}
& Appeal to nature $>$ Appeal to authority & $<.01$ \\
& Appeal to nature $>$ Appeal to majority & $<.001$ \\
& Appeal to tradition $>$ Appeal to authority & $<.001$ \\
& Appeal to tradition $>$ Appeal to majority & $<.001$ \\
& Appeal to worse problems $<$ Appeal to nature & $<.001$ \\
& Appeal to worse problems $<$ Appeal to tradition & $<.001$ \\
& False dilemma $<$ Appeal to nature & $<.001$ \\
& False dilemma $<$ Appeal to tradition & $<.001$ \\
& Hasty generalization $<$ Appeal to tradition & $<.01$ \\
& Hasty generalization $<$ Appeal to worse problems & $<.01$ \\
& Hasty generalization $<$ False dilemma & $<.01$ \\
& Slippery slope $<$ Appeal to tradition & $<.001$ \\

\bottomrule
\end{tabular}
\caption{Pairwise comparisons across logical fallacy types for confidence ratings. Only selected statistically significant comparisons are shown for readability.}
\label{tab:appendix_confidence_fallacy_type_pairwise}
\end{table*}

Humans and LLMs varied in confidence across logical fallacy types (Table \ref{tab:appendix_confidence_fallacy_type_pairwise}; Welch two-sample $t$-tests with Holm correction were used due to unequal variances across groups). Humans reported higher confidence for \textit{slippery slope}, \textit{appeal to worse problems}, and \textit{appeal to majority} ($p_{adj}<.01$), whereas confidence was comparatively lower for \textit{hasty generalization}. For LLMs, confidence was highest for \textit{appeal to tradition} and \textit{appeal to nature}, both rated significantly higher than several other fallacies ($p_{adj}<.001$). In contrast, \textit{hasty generalization}, \textit{false dilemma}, and \textit{appeal to worse problems} generally received comparatively lower confidence ratings.

\subsubsection{Human and LLMs differ in Confidence Calibration} \label{appendix: human_and_llms_differ_in_confidence_calibration}

Interestingly, humans and LLMs showed different relationships between confidence and susceptibility across fallacy types. Human evaluators were often more susceptible to certain fallacies, such as \textit{hasty generalization}, despite expressing relatively lower confidence for them, whereas higher confidence for \textit{slippery slope} and \textit{appeal to majority} coincided with lower perceived logical accuracy ratings. One possible interpretation is that human evaluators may sometimes assign relatively high logical accuracy ratings to fallacious arguments even when they are uncertain about their judgments. For example, the comparatively high logical accuracy ratings but lower confidence ratings for \textit{hasty generalization} suggest that participants may have sensed ambiguity or difficulty detecting the reasoning flaw while still evaluating the argument favorably overall. 

This pattern is also consistent with prior work on hallucination evaluation, where participants assigned higher perceived accuracy ratings to minor hallucinations despite expressing lower confidence in those judgments \cite{nahar2025catch}. Perhaps, human evaluators may sometimes remain susceptible to logically flawed content in cases where they are uncertain. In contrast, LLM confidence patterns more closely aligned with their logical accuracy judgments: fallacy types receiving higher logical accuracy ratings, such as \textit{appeal to nature} and \textit{appeal to tradition}, also tended to receive higher confidence ratings. Together, these patterns suggest that humans and LLMs may differ in how confidence calibration relates to reasoning susceptibility across fallacy types.



\subsection{Results: Trust and Evaluation} \label{appendix: trust and evaluation results}
Both the trust scales (Cronbach's $\alpha=.84$) and the overall evaluation scales (Cronbach's $\alpha=.89$) demonstrated high internal consistency. Therefore, items were averaged into composite measures. Descriptive statistics showed a similar pattern for both trust and overall evaluation ratings across conditions. For trust, the Human condition received the highest ratings ($M=3.69$, 95\% CI [3.52, 3.86]), followed by Human+AI ($M=3.39$, 95\% CI [3.21, 3.56]), AI+Human ($M=3.06$, 95\% CI [2.88, 3.23]), AI ($M=2.93$, 95\% CI [2.76, 3.10]), and Control ($M=2.83$, 95\% CI [2.65, 3.01]). A similar pattern emerged for overall evaluation ratings: Human received the highest evaluations ($M=3.91$, 95\% CI [3.75, 4.07]), followed by Human+AI ($M=3.63$, 95\% CI [3.47, 3.80]), AI+Human ($M=3.23$, 95\% CI [3.07, 3.39]), AI ($M=3.15$, 95\% CI [2.99, 3.31]), and Control ($M=3.11$, 95\% CI [2.95, 3.28]).  

Overall, conditions associated with human involvement received higher trust and evaluation ratings than fully AI-generated or undisclosed-source conditions. The post-hoc pairwise comparisons for trust and overall evaluation scores, using Tukey-adjusted $p$-values, are presented in Table~\ref{tab:appendix_rq2_pairwise}. In addition, trust and overall evaluation scores were strongly correlated ($r = .89$, $p < .001$), indicating that participants who trusted the commenter more also evaluated the commenter more positively.

\begin{table*}[t]
\centering
\small
\begin{tabular}{llcccc}
\toprule
\textbf{Dependent Variable} & \textbf{Comparison} & \textbf{Mean Diff.} & \textbf{SE} & \textbf{$t$} & \textbf{$p_{adj}$} \\
\midrule

\multirow{10}{*}{Trust}
& Control vs Human & $-0.860$ & 0.126 & $-6.83$ & $<.001$ \\
& Control vs AI & $-0.101$ & 0.126 & $-0.80$ & $.929$ \\
& Control vs Human+AI & $-0.557$ & 0.127 & $-4.39$ & $<.001$ \\
& Control vs AI+Human & $-0.229$ & 0.127 & $-1.81$ & $.371$ \\
& Human vs AI & $0.759$ & 0.124 & $6.10$ & $<.001$ \\
& Human vs Human+AI & $0.303$ & 0.125 & $2.42$ & $.112$ \\
& Human vs AI+Human & $0.632$ & 0.125 & $5.05$ & $<.001$ \\
& AI vs Human+AI & $-0.456$ & 0.125 & $-3.64$ & $.003$ \\
& AI vs AI+Human & $-0.127$ & 0.125 & $-1.02$ & $.847$ \\
& Human+AI vs AI+Human & $0.329$ & 0.126 & $2.61$ & $.070$ \\

\midrule

\multirow{10}{*}{Overall Evaluation}
& Control vs Human & $-0.797$ & 0.118 & $-6.76$ & $<.001$ \\
& Control vs AI & $-0.038$ & 0.118 & $-0.33$ & $.998$ \\
& Control vs Human+AI & $-0.521$ & 0.119 & $-4.39$ & $<.001$ \\
& Control vs AI+Human & $-0.118$ & 0.118 & $-0.99$ & $.859$ \\
& Human vs AI & $0.758$ & 0.116 & $6.52$ & $<.001$ \\
& Human vs Human+AI & $0.276$ & 0.117 & $2.35$ & $.130$ \\
& Human vs AI+Human & $0.679$ & 0.117 & $5.81$ & $<.001$ \\
& AI vs Human+AI & $-0.483$ & 0.117 & $-4.12$ & $<.001$ \\
& AI vs AI+Human & $-0.079$ & 0.117 & $-0.68$ & $.961$ \\
& Human+AI vs AI+Human & $0.403$ & 0.118 & $3.42$ & $.006$ \\

\bottomrule
\end{tabular}
\caption{Pairwise comparisons for trust and overall evaluation ratings across source conditions (Tukey-adjusted).}
\label{tab:appendix_rq2_pairwise}
\end{table*}

\subsection{Post-Session Results} \label{appendix: post_session_results}

\subsubsection{Manipulation Check} Participants generally identified the assigned source conditions correctly, with an overall manipulation-check accuracy of 82.2\%. Accuracy was higher for the Human (87.4\%), Human+AI (88.0\%), AI (86.4\%), and no source disclosure (81.6\%) conditions, suggesting that participants largely perceived the intended source manipulations. Accuracy was lowest for the AI with human assistance (67.3\%) condition. Some participants identified the comments as written by human with AI assistance, indicating that those participants may have perceived these two forms of collaboration as conceptually similar or difficult to distinguish. Nevertheless, participants were generally able to articulate nuanced reasoning about authorship, human oversight, and AI involvement when explaining their interpretations of the comment writing process and their source preferences.

\subsubsection{Computer and AI Expertise} \label{appendix: computer and AI expertise}
The results of computer expertise, AI expertise, and AI usage frequency are presented in Table \ref{tab:ai_expertise}. Notably, the participant population reported relatively high levels of computer familiarity, AI expertise, and AI usage frequency. Despite this, participants remained susceptible to source-label bias, suggesting that \textbf{familiarity with and higher usage of AI systems alone may not protect users from heuristic reasoning based on perceived authorship or credibility cues}. \textit{Pairwise comparisons showed no meaningful differences in these measures across experimental conditions (all $ps$ > .05), indicating that the observed effects were unlikely to be driven by variations in participants’ technical or AI-related backgrounds. Therefore, these variables were not included as covariates in the main analyses.}

\begin{table}[hbt!]
\centering
\small
\begin{tabular}{p{5.9cm}p{1.2cm}}
\toprule
\textbf{Measure} & \textbf{N} \\
\midrule

\multicolumn{2}{l}{\textbf{Computer Expertise}} \\
Expert (5) & 78 \\
Advanced (4) & 243 \\
Intermediate (3) & 156 \\
Basic (2) & 26 \\
Novice (1) & 2 \\

\midrule
\multicolumn{2}{l}{\textbf{AI Expertise}} \\
Expert (5) & 21 \\
Advanced (4) & 197 \\
Intermediate (3) & 207 \\
Basic understanding (2) & 72 \\
No experience (1) & 8 \\

\midrule
\multicolumn{2}{l}{\textbf{AI Usage Frequency}} \\
Very frequently (5) & 149 \\
Frequently (4) & 182 \\
Occasionally (3) & 120 \\
Rarely (2) & 36 \\
Never (1) & 18 \\

\bottomrule
\end{tabular}
\caption{Participants' computer and AI expertise characteristics ($N=505$).}
\label{tab:ai_expertise}
\end{table}

\subsubsection{News Source Preference and Political Alignment} \label{appendix_post-session_news_source}

Participants' news source preferences are presented in Table \ref{tab:news_source}. Importantly, \textbf{56\% of participants reported relying on social media platforms for news consumption}, environments that are increasingly populated by AI-generated and AI-assisted content, as evident by platforms such as Instagram, which are investigating AI-generated comments to enhance user engagement \cite{marketingtechnews}, while researchers have employed AI bots on Reddit to test commenters’ perceptions \cite{theverge}. Thus, \textbf{source labels have a crucial impact on how these participants evaluate and perceive the sources from which they consume news}. As our participants also exhibited source bias, these findings raise concerns that AI-generated content paired with signals of human involvement may become especially persuasive in online ecosystems where human and AI-generated communication are increasingly difficult to distinguish \cite{spitale2023ai}.

\begin{table}[t]
\centering
\small
\begin{tabular}{{p{5.8cm}p{1.2cm}}}
\toprule
\textbf{Preferred News Source} & \textbf{N} \\
\midrule
Social media platforms & 282 \\
News websites and apps & 271 \\
Television news channels & 188 \\
Search engines & 166 \\
Podcasts or radio & 122 \\
Friends, family, or word of mouth & 97\\
Newspapers & 85\\
Government websites & 35\\
Others, please specify & 13\\
\bottomrule
\end{tabular}
\caption{Participants' preferred news sources. Note. Participants were allowed to select multiple options.}
\label{tab:news_source}
\end{table}

\begin{table}[hbt!]
\centering
\small
\begin{tabular}{p{5.8cm}p{1.2cm}}
\toprule
\textbf{Political Alignment} & \textbf{N} \\
\midrule
Completely left (1) & 123 \\
Somewhat left (2) & 122 \\
Neutral; neither left nor right (3) & 78 \\
Somewhat right (4) & 116 \\
Completely right (5) & 66 \\
\bottomrule
\end{tabular}
\caption{Participants' self-reported political alignment ($N=505$).}
\label{tab:political_alignment}
\end{table}

\begin{table*}[t]
\centering
\small
\begin{tabular}{lcccc}
\toprule
\textbf{Source} & \textbf{Mean Rank} & \textbf{Median} & \textbf{\#1 Rank} & \textbf{\#5 Rank} \\
\midrule
Human & 1.57 & 1 & 348 & 12 \\
Human+AI & 2.28 & 2 & 60 & 13 \\
AI+Human & 3.00 & 3 & 47 & 11 \\
AI & 3.31 & 4 & 48 & 8 \\
No disclosure & 4.83 & 5 & 2 & 461 \\
\midrule
\multicolumn{5}{l}{\textbf{Friedman Test:} $\chi^2(4)=1214.20,\ p<.001$} \\
\midrule
\multicolumn{5}{l}{\textbf{Significant Pairwise Comparisons (Wilcoxon signed-rank, Holm corrected)}} \\
\midrule
Human $>$ Human+AI ($p<.001$) \\
Human $>$ AI+Human ($p<.001$) \\
Human $>$ AI ($p<.001$) \\
Human $>$ No disclosure ($p<.001$) \\
Human+AI $>$ AI+Human ($p<.001$) \\
Human+AI $>$ AI ($p<.001$) \\
Human+AI $>$ No disclosure ($p<.001$) \\
AI+Human $>$ AI ($p<.001$) \\
AI+Human $>$ No disclosure ($p<.001$) \\
AI $>$ No disclosure ($p<.001$) \\
\bottomrule
\end{tabular}
\caption{Preference rankings for comment source types. Lower ranks indicate greater preference. Participants strongly preferred Human and Human+AI sources, whereas No Disclosure was overwhelmingly least preferred.}
\label{tab:source_preferences}
\end{table*}

Participants' political alignment results are presented in Table \ref{tab:political_alignment}. Participants represented a politically diverse sample, and political alignment results were similar across conditions (all $ps > .05$). Prior research on misinformation often reports ideological asymmetries and politically motivated reasoning in how individuals evaluate information \cite{pennycook2021psychology}. In contrast, our study examined logical reasoning judgments in relatively politically neutral contexts rather than explicit misinformation settings. The presence of source-label effects across a politically diverse participant pool, therefore, suggests that heuristic reliance on perceived authorship may reflect a broader cognitive tendency that extends beyond specific ideological groups. It is also possible that participants perceived our stimuli as relatively politically neutral, as intended in our study design, which reduced the influence of partisan identity on reasoning judgments. \textit{Neither news source preference nor political alignment showed any meaningful differences across experimental conditions (all $ps$ > .05). Therefore, these variables were not included as covariates in the main analyses.}

\subsubsection{Participants' Interpretation of the Comment Writing Process} \label{appendix_post-session_comment_writing_process}

To better understand how participants perceived the writing process behind each comment source, particularly the two AI-assistance conditions (human with AI assistance and AI with human assistance) and how participants distinguished between them, we conducted an exploratory descriptive analysis of open-ended explanations. As this analysis was intended to contextualize the quantitative preference results rather than serve as a full qualitative study, we used a simplified form of thematic coding \cite{gibbs2007thematic}. One author inductively identified recurring themes from an initial review of the responses and then applied these themes to code the remaining data. Themes were not mutually exclusive, as a single response could contain multiple rationales.

The responses reveal that participants did not merely repeat the assigned labels; rather, they actively interpreted the comments through perceived cues of authorship, authenticity, grammar, emotional tone, and reasoning style. Across conditions, participants frequently relied on heuristic cues such as typos, conversational tone, emotional language, grammatical inconsistency, and perceived \lq\lq genericness\rq\rq \space to infer whether comments were written by humans or AI. Many participants associated human-authored comments with emotional nuance, lived experience, grammatical imperfections, and authentic social-media-like behavior, whereas AI-authored comments were often described as overly polished, generic, repetitive, emotionally flat, or structurally uniform. Interestingly, some participants interpreted poor grammar or illogical reasoning as evidence of human authorship, while others interpreted highly structured or overly confident writing as evidence of AI generation. 

The human with AI assistance and AI with human assistance conditions revealed especially important differences in how participants conceptualized collaboration between humans and AI. \textbf{In the human with AI assistance condition, participants frequently described the comments as primarily human-authored but refined, polished, or edited using AI tools.} AI was commonly framed as a supportive assistant that improved clarity, organization, grammar, or flow while leaving humans in control of the core ideas and opinions. For example, one participant described the comments as \textit{\lq\lq human opinions that have been lightly polished,\rq\rq} while another suggested that \textit{\lq\lq people wrote out the comments but used AI to help them write it out\rq\rq}. Participants often interpreted this condition as preserving human agency, authenticity, and accountability despite AI involvement.

In contrast, \textbf{participants in the AI with human assistance condition more frequently interpreted the comments as primarily AI-generated, with humans serving only supervisory, corrective, or editorial roles.} Responses commonly described humans as \lq\lq giving feedback,\rq\rq \space \lq\lq checking,\rq\rq \space or \lq\lq guiding\rq\rq \space AI outputs rather than originating the reasoning itself. Participants frequently framed the AI as the primary author and humans as secondary overseers. For example, participants described the comments as \textit{\lq\lq AI wrote something and a human gave feedback,\rq\rq} \space or \textit{\lq\lq the AI wrote the comments which were later evaluated by human.\rq\rq}\space This distinction suggests that participants were sensitive not only to the presence of AI, but also to perceived authorship hierarchy and human control within collaborative systems.

Interestingly, even when explicit source labels were provided, some participants continued to disagree with or reinterpret the assigned condition based on stylistic expectations. Several participants in the human condition suspected AI involvement because the comments appeared too polished, generic, or structurally consistent, whereas some participants in the AI condition inferred human authorship due to typos, emotional language, or conversational tone. This suggests that participants do not passively accept source labels; rather, they combine disclosure cues with their own folk theories about how humans and AI \textit{should} write. These results point to an interesting insight: source perception is shaped not only by explicit labels but also by users’ evolving expectations, heuristics, and assumptions about human-AI interaction.

\subsubsection{Participants' Source Preference Ranks and Their Rationales} \label{appendix_post-session_source_preferences}

All participants were asked to rank the five sources in descending order of preference. \textbf{Across conditions, participants claimed to strongly prefer comments labeled as written by human or written by human with AI assistance}, whereas comments without source disclosure were overwhelmingly disfavored. This pattern mirrors the trust and logical reasoning results (Table \ref{tab:source_preferences}), suggesting that source labels influence not only reasoning judgments but also explicit source preferences. Interestingly, \textbf{participants strongly disliked the absence of source information}, indicating that users may actively seek authorship cues even when such cues can bias evaluation. Preference rankings for source types did not significantly differ across experimental conditions, suggesting that the assigned condition did not affect participants' explicit source preferences. 

The strong preference for particular labels, combined with increased susceptibility to flawed reasoning under those conditions, raises concerns that AI-generated content paired with signals of human involvement may become especially persuasive in AI-mediated online environments. At the same time, participants’ strong dislike of missing source information highlights a tension for platform design: while users actively seek authorship cues, those same cues may unintentionally bias evaluation.

\textbf{Participants' rationales behind their source preferences.} To better understand participants' source preference rankings, we conducted an exploratory descriptive analysis of the open-ended explanations provided after the ranking task. As this analysis was intended to contextualize the quantitative preference results rather than serve as a full qualitative study, we used a simplified form of thematic coding \cite{gibbs2007thematic}. One author inductively identified recurring themes from an initial review of the responses, then applied these themes to code the remaining data. Themes are not mutually exclusive, as a single response may contain multiple rationales.


Overall, participants' explanations aligned closely with the ranking results. Many participants preferred human-authored comments because they perceived them as more authentic, emotionally grounded, nuanced, and reflective of lived experience. In contrast, AI-generated comments were frequently described as less trustworthy, less genuine, or prone to errors and hallucinations. Participants also expressed strong preferences for transparency and source disclosure, often describing undisclosed sources as suspicious, difficult to evaluate, or lacking accountability. These responses help explain why human and human with AI assistance were preferred most, whereas no source disclosure was overwhelmingly ranked last. More broadly, the qualitative responses reinforce our main finding that source labels shape not only reasoning judgments, but also explicit preferences and perceived credibility.

As evident in participants' interpretations of the comment writing process, they appeared to interpret human with AI assistance and AI with human assistance as fundamentally different forms of collaboration. Responses frequently described human with AI assistance as preserving human agency and accountability while using AI as a supportive tool for clarity, organization, or efficiency. As one participant noted: \textit{\lq\lq Human comments are genuine. Some AI assistance can be used for putting thoughts into words, which is OK. The comments should not be written by AI or mostly AI. That is not genuine\rq\rq}. Several participants similarly viewed human with AI assistance as primarily human-authored content that had been refined or edited using AI assistance (\textit{\lq\lq I don't mind if a real human uses AI for flow, editing or slight enhancements\rq\rq}). In contrast, AI with human assistance was often perceived as primarily AI-driven, with humans serving only supervisory or corrective roles. One participant described it as: \textit{\lq\lq The primary reasoning and output comes from AI, with a person merely serving to steer it one way\rq\rq}. These responses suggest that participants were sensitive not only to the presence of AI, but also to perceived authorship hierarchy and human control within AI-assisted collaboration.

\subsection{Robustness Across Prompting Strategies} \label{appendix: robust analysis across prompting strategies}
To evaluate robustness across prompting strategies, we compared the baseline prompting setting with chain-of-thought prompting \cite{wei2022chain} and expert-evaluator framing \cite{xu2023expertprompting}. These prompting strategies were applied only to LLMs, as our goal was to assess the robustness of model-based evaluation behavior rather than to manipulate human reasoning processes.

To facilitate direct comparisons across prompting strategies, we used the same rating scales for perceived logical accuracy and confidence across all settings. \textbf{The robustness analyses indicate that LLMs remained broadly stable across source labels, consistent with our main findings.}

\subsubsection{Robustness Prompts}

\textbf{Chain-of-thought:} Chain-of-thought (CoT) is a series of intermediate reasoning steps, that can significantly improve the ability of large language models to perform complex reasoning \cite{wei2022chain}. For CoT, we used the following prompt.

\vspace{0.1in}
\textit{[If source disclosed] All comments are written by a [human/AI/human with AI assistance/AI with human assistance].}

\textit{Analyze the comment step by step: 1. Identify the main claim. 2. Determine whether the reasoning contains logical flaws or fallacies. 3. Briefly explain your reasoning.}

\textit{Then answer the following two questions.}

\textit{Rate how logically accurate you find the reasoning in the comment. Here, 1 = not at all logically accurate, 2= slightly logically accurate, 3 = somewhat logically accurate, 4 = mostly logically accurate, and 5 = completely logically accurate.}

\textit{How confident are you in your assessment of the comment? Here, 1 = not at all confident, 2 = slightly confident, 3 = somewhat confident, 4 = mostly confident, and 5 = completely confident.} 

\vspace{0.1in}
\textbf{Expert evaluator framing:} Expert evaluator framing is a prompting technique that improves judgment quality by assigning the model a specific professional evaluator role, that can improve response quality \cite{xu2023expertprompting}. For expert framing, we utilized the following prompt:

\vspace{0.1in}
\textit{You are an expert in logical reasoning and argument analysis. Carefully evaluate the logical quality of the following comment.}

\textit{[If source disclosed] All comments are written by a [human/AI/human with AI assistance/AI with human assistance].}

\textit{Answer the following two questions.}

\textit{Rate how logically accurate you find the reasoning in the comment. Here, 1 = not at all logically accurate, 2= slightly logically accurate, 3 = somewhat logically accurate, 4 = mostly logically accurate, and 5 = completely logically accurate.}

\textit{How confident are you in your assessment of the comment? Here, 1 = not at all confident, 2 = slightly confident, 3 = somewhat confident, 4 = mostly confident, and 5 = completely confident.} 

\subsubsection{Perceived Logical Accuracy Across Prompting Strategies}
\begin{figure}[hbt!]
\centering
  \includegraphics[width=\columnwidth]{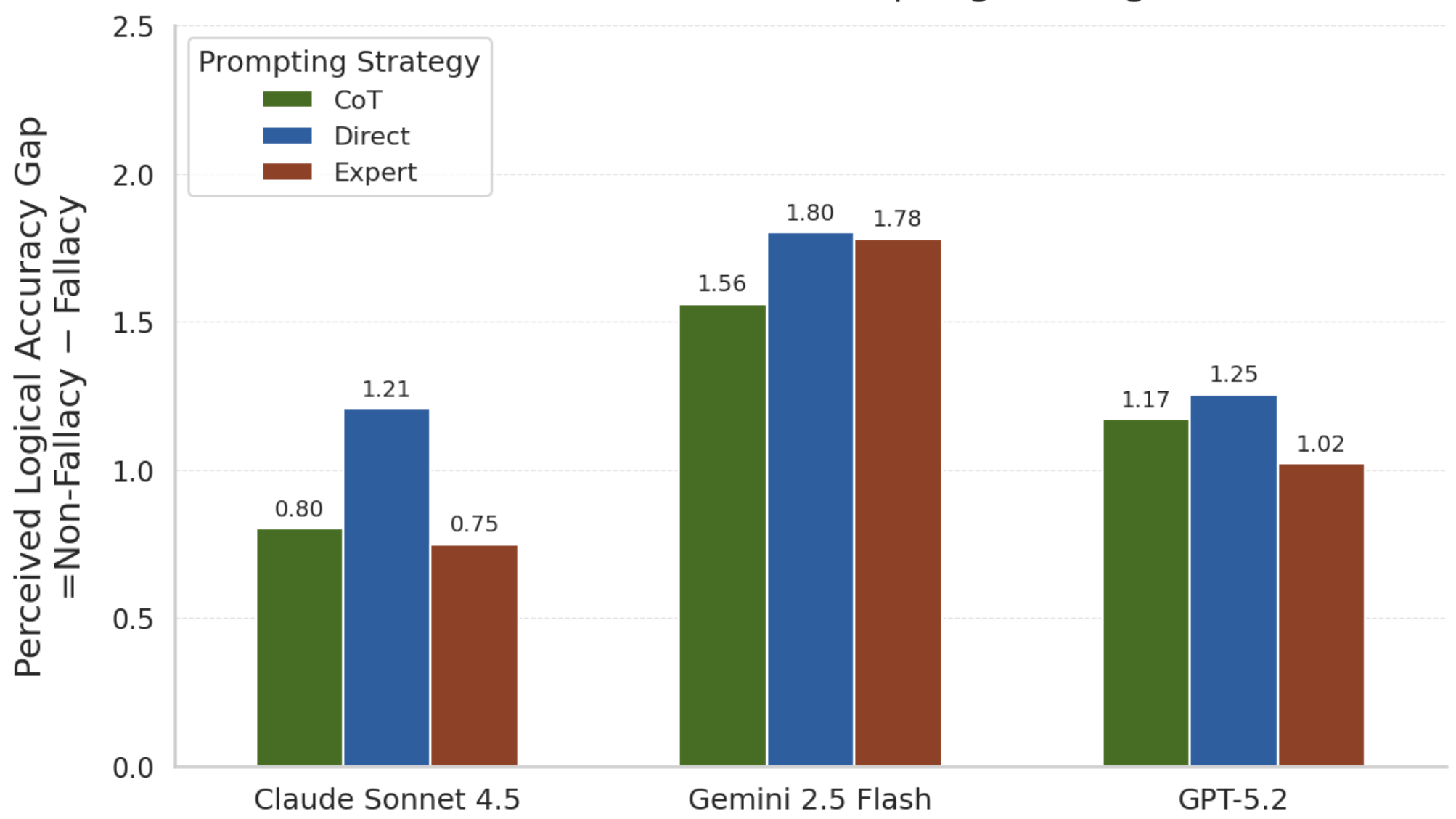}
  \caption{Perceived logical accuracy gap = Non-fallacy - Fallacy, by model and prompting strategy. Higher values are better, as they indicate greater sensitivity to fallacious content.}
  \label{fig:appendix_robustness_gap}
\end{figure}

To evaluate the robustness of our findings, we repeated the LLM evaluations using chain-of-thought (CoT) and expert-style prompting strategies in addition to direct prompting, as shown in Table \ref{tab:prompting_robustness_means_pa}. Overall, the primary qualitative findings remained stable across prompting methods. Across all models, prompts, and source conditions, non-fallacious comments consistently received significantly higher logical accuracy ratings than fallacious comments, indicating robust sensitivity to reasoning quality. As illustrated in Figure~\ref{fig:appendix_robustness_gap}, this fallacy-non-fallacy gap remained consistently positive across all models and prompting strategies. At the same time, source-label effects were largely absent: across all model, prompt, and fallacy combinations, only one significant across-source comparison emerged, for Claude Sonnet 4.5 under chain-of-thought prompting in the non-fallacy condition, where Control ($M = 3.21$) exceeded AI ($M = 2.94$, $p = .01$).

\begin{table}[t]
\centering
\small
\setlength{\tabcolsep}{4pt}
\begin{tabular}{llcccc}
\toprule
\textbf{Model} & \textbf{\shortstack[l]{Source\\Condition}} & \textbf{Fallacy} & \textbf{Direct} & \textbf{CoT} & \textbf{Expert} \\
\midrule

\multirow{10}{*}{GPT-5.2}
& Control & Present & $2.44$ & $2.29$ & $2.17$ \\
& Control & Absent & $3.79$ & $3.54$ & $3.31$ \\
& Human & Present & $2.54$ & $2.35$ & $2.19$ \\
& Human & Absent & $3.77$ & $3.56$ & $3.29$ \\
& AI & Present & $2.29$ & $2.17$ & $2.02$ \\
& AI & Absent & $3.50$ & $3.29$ & $3.02$ \\
& Human+AI & Present & $2.38$ & $2.29$ & $2.17$ \\
& Human+AI & Absent & $3.65$ & $3.46$ & $3.10$ \\
& AI+Human & Present & $2.33$ & $2.15$ & $2.04$ \\
& AI+Human & Absent & $3.54$ & $3.25$ & $2.98$ \\

\midrule

\multirow{10}{*}{\shortstack[l]{Claude\\Sonnet 4.5}}
& Control & Present & $1.92$ & $1.81$ & $1.83$ \\
& Control & Absent & $3.21$ & $2.92$ & $2.71$ \\
& Human & Present & $1.94$ & $1.87$ & $1.90$ \\
& Human & Absent & $3.27$ & $2.67$ & $2.69$ \\
& AI & Present & $1.90$ & $1.75$ & $1.77$ \\
& AI & Absent & $2.94$ & $2.35$ & $2.46$ \\
& Human+AI & Present & $1.92$ & $1.78$ & $1.85$ \\
& Human+AI & Absent & $3.15$ & $2.52$ & $2.50$ \\
& AI+Human & Present & $1.88$ & $1.77$ & $1.75$ \\
& AI+Human & Absent & $3.02$ & $2.53$ & $2.48$ \\

\midrule

\multirow{10}{*}{\shortstack[l]{Gemini\\2.5 Flash}}
& Control & Present & $2.04$ & $1.65$ & $2.06$ \\
& Control & Absent & $3.94$ & $3.29$ & $3.96$ \\
& Human & Present & $2.12$ & $1.58$ & $2.10$ \\
& Human & Absent & $3.94$ & $3.10$ & $3.88$ \\
& AI & Present & $2.00$ & $1.44$ & $1.98$ \\
& AI & Absent & $3.73$ & $3.02$ & $3.77$ \\
& Human+AI & Present & $2.15$ & $1.52$ & $1.98$ \\
& Human+AI & Absent & $3.92$ & $3.06$ & $3.79$ \\
& AI+Human & Present & $2.00$ & $1.52$ & $1.94$ \\
& AI+Human & Absent & $3.79$ & $3.04$ & $3.56$ \\

\bottomrule
\end{tabular}
\caption{Mean perceived logical accuracy ratings across source conditions, fallacy presence, and prompting strategies for LLMs.}
\label{tab:prompting_robustness_means_pa}
\end{table}

Prompting strategies primarily affected evaluation strictness rather than the broader pattern of judgments. CoT and expert prompting frequently produced lower logical accuracy ratings than direct prompting, particularly for non-fallacious comments, suggesting more conservative or cautious evaluation behavior. This pattern was especially pronounced for Claude Sonnet 4.5 and Gemini 2.5 Flash, which showed larger reductions under deliberative prompting, whereas GPT-5.2 exhibited more consistent decreases under expert prompting.

Importantly, despite these shifts in absolute ratings, the relative structure of evaluations remained largely unchanged across prompting methods, suggesting that prompting altered calibration more strongly than underlying reasoning preferences or source-label sensitivity. These findings further support the conclusion that source-label susceptibility was primarily a human vulnerability rather than a property of LLMs.

\subsubsection{Confidence Ratings Across Prompting Strategies}

\begin{table}[t]
\centering
\small
\setlength{\tabcolsep}{4pt}
\begin{tabular}{llcccc}
\toprule
\textbf{Model} & \textbf{\shortstack[l]{Source\\Condition}} & \textbf{Fallacy} & \textbf{Direct} & \textbf{CoT} & \textbf{Expert} \\
\midrule

\multirow{10}{*}{GPT-5.2}
& Control & Present & $4.29$ & $4.29$ & $4.31$ \\
& Control & Absent & $4.33$ & $4.12$ & $4.12$ \\
& Human & Present & $4.29$ & $4.27$ & $4.29$ \\
& Human & Absent & $4.42$ & $4.08$ & $4.17$ \\
& AI & Present & $4.46$ & $4.33$ & $4.40$ \\
& AI & Absent & $4.42$ & $4.10$ & $4.19$ \\
& Human+AI & Present & $4.38$ & $4.33$ & $4.33$ \\
& Human+AI & Absent & $4.40$ & $4.08$ & $4.19$ \\
& AI+Human & Present & $4.46$ & $4.40$ & $4.46$ \\
& AI+Human & Absent & $4.48$ & $4.17$ & $4.25$ \\

\midrule

\multirow{10}{*}{\shortstack[l]{Claude\\Sonnet 4.5}}
& Control & Present & $4.27$ & $4.60$ & $4.29$ \\
& Control & Absent & $4.04$ & $4.10$ & $4.02$ \\
& Human & Present & $4.25$ & $4.57$ & $4.33$ \\
& Human & Absent & $4.02$ & $4.08$ & $4.08$ \\
& AI & Present & $4.29$ & $4.71$ & $4.35$ \\
& AI & Absent & $4.06$ & $4.17$ & $4.04$ \\
& Human+AI & Present & $4.29$ & $4.59$ & $4.31$ \\
& Human+AI & Absent & $4.04$ & $4.08$ & $4.04$ \\
& AI+Human & Present & $4.31$ & $4.67$ & $4.38$ \\
& AI+Human & Absent & $4.06$ & $4.13$ & $4.04$ \\

\midrule

\multirow{10}{*}{\shortstack[l]{Gemini\\2.5 Flash}}
& Control & Present & $4.67$ & $4.92$ & $4.85$ \\
& Control & Absent & $4.71$ & $4.79$ & $4.77$ \\
& Human & Present & $4.60$ & $4.96$ & $4.85$ \\
& Human & Absent & $4.65$ & $4.88$ & $4.83$ \\
& AI & Present & $4.67$ & $4.98$ & $4.83$ \\
& AI & Absent & $4.71$ & $4.88$ & $4.81$ \\
& Human+AI & Present & $4.67$ & $4.94$ & $4.88$ \\
& Human+AI & Absent & $4.50$ & $4.90$ & $4.79$ \\
& AI+Human & Present & $4.67$ & $4.98$ & $4.88$ \\
& AI+Human & Absent & $4.56$ & $4.94$ & $4.85$ \\

\bottomrule
\end{tabular}
\caption{Mean confidence ratings across source conditions, fallacy presence, and prompting strategies for LLMs.}
\label{tab:prompting_robustness_means_confidence}
\end{table}

Across all three models, no significant differences emerged across source conditions for confidence ratings (Table \ref{tab:prompting_robustness_means_confidence}), indicating that LLM confidence judgments remained largely source-agnostic across prompting strategies. This mirrors the confidence findings from the main LLM evaluation and further supports the conclusion that source-label differences were minimal across prompting strategies. At the same time, prompting strategies produced shifts in absolute values of confidence rather than source sensitivity.

Claude Sonnet 4.5 exhibited the strongest prompting-related effects. Under chain-of-thought (CoT) prompting, Claude frequently assigned higher confidence ratings than under direct or expert prompting, particularly for fallacious comments. Interestingly, these increases in confidence occurred even though CoT prompting simultaneously lowered perceived logical-accuracy ratings in the primary robustness analyses, suggesting more decisive identification of flawed reasoning under deliberative prompting. Claude also showed lower confidence ratings for non-fallacious comments relative to fallacious comments across several prompting settings. Gemini 2.5 Flash showed comparatively stable confidence ratings across prompting strategies, despite maintaining consistently high overall confidence levels. GPT-5.2 exhibited modest decreases in confidence under CoT and expert prompting relative to direct prompting, particularly for non-fallacious comments.

Importantly, despite these prompt-related shifts in confidence calibration, the absence of systematic source-label effects remained stable across models and prompting strategies. Overall, the robustness analyses suggest that prompting strategies influenced confidence calibration more strongly than source-label sensitivity, reinforcing our central conclusion that source-label susceptibility was primarily characteristic of human evaluators rather than LLMs.

\subsection{Robustness Analysis Based on Source-Label Identification} \label{appendix: manipulation-check analysis}
As participants had lower accuracy in identifying the AI+Human condition in the manipulation check (67.3\%), we conducted additional robustness analyses restricted to participants who correctly identified their assigned source label. This restriction reduced the analytic sample from the full dataset to $N=415$ (Control: 80, Human: 90, AI: 89, Human+AI: 88, AI+Human: 68). We repeated the main analyses on this restricted sample to assess whether the observed source-label effects, particularly differences between the Human+AI and AI+Human conditions, were robust to participants’ recognition of the assigned authorship label. Overall, the results for perceived logical accuracy, confidence, and trust remained consistent, whereas the overall-evaluation results showed some differences in the restricted sample.

\subsubsection{Perceived Logical Accuracy}
For perceived logical accuracy, restricting the analysis to participants who correctly identified their assigned source label ($N=415$) produced results that were consistent with the main analysis. We observed a significant Condition $\times$ Fallacy interaction, $F(4,410)=87.84$, $p<.001$. For fallacies, perceived logical accuracy was significantly higher in the Human ($M=3.35$) and Human+AI ($M=3.35$) conditions than in the Control ($M=2.35$), AI ($M=2.48$), and AI+Human ($M=2.59$) conditions (all $ps<.001$), while Human and Human+AI did not differ from each other. In contrast, perceived logical accuracy for non-fallacious comments did not differ across any source conditions (all $ps>.5$). Moreover, fallacious comments were rated significantly less logically accurate than non-fallacious comments within every condition (all $ps<.001$), with substantially smaller fallacy penalties in the Human and Human+AI conditions.

\subsubsection{Confidence}
The results for confidence were also consistent. There was no significant effect of source condition, $F(4,410)=1.12$, $p=.347$, while confidence was significantly lower for fallacies vs. non-fallacies, $F(1,410)=8.36$, $p=.004$; the Condition $\times$ Fallacy interaction was not significant, $F(4,410)=1.90$, $p=.109$. Pairwise comparisons showed no significant differences in confidence across source conditions for either fallacious or non-fallacious comments. Consistent with the main analysis, the fallacy-related decrease in confidence was significant only in the Control condition (NonFallacy: $M=3.90$; Fallacy: $M=3.79$; $p=.002$) and was not significant in any of the labeled source conditions (all $ps>.14$).

\subsubsection{Trust}
For trust, restricting the analysis to participants who correctly identified their assigned source label ($N=415$) yielded a significant effect of condition, $F(4,410)=17.28$, $p<.001$. Human received the highest trust ratings ($M=3.69$, $SE=.09$), followed by Human+AI ($M=3.34$, $SE=.09$), AI+Human ($M=2.98$, $SE=.11$), AI ($M=2.88$, $SE=.09$), and Control ($M=2.71$, $SE=.10$). Consistent with the main analysis, Tukey-adjusted comparisons showed that Human received significantly higher trust ratings than Control, AI, and AI+Human (all $ps<.001$), while Human+AI received significantly higher ratings than Control ($p<.001$) and AI ($p=.005$). No other pairwise differences were significant.

\subsubsection{Overall Evaluation}
For overall evaluation, restricting the analysis to participants who correctly identified their assigned source label ($N=415$) yielded a significant effect of condition, $F(4,410)=16.08$, $p<.001$. Human received the highest evaluations ($M=3.90$, $SE=.09$), followed by Human+AI ($M=3.57$, $SE=.09$), AI+Human ($M=3.22$, $SE=.10$), AI ($M=3.14$, $SE=.09$), and Control ($M=3.02$, $SE=.09$). Tukey-adjusted comparisons showed that Human was rated significantly higher than Control, AI, and AI+Human (all $ps<.001$), but did not significantly differ from Human+AI ($p=.062$). Human+AI was rated significantly higher than Control ($p<.001$) and AI ($p=.006$); however, its advantage over AI+Human was only suggestive and did not reach statistical significance ($p=.072$). No other pairwise differences were significant.

\subsection{Additional Robustness Analyses} \label{appendix: additional robustness analyses}

\subsubsection{Covariate Analyses}
In the initial analysis, participants’ AI expertise, computer expertise, political alignment, and news-source preferences were checked for balance across conditions, but we did not include them as covariates. As additional robustness analysis, we reran the LMER models with AI expertise, computer expertise, political alignment, and news-source preference as covariates. The conclusions remained unchanged, with a significant Condition × Fallacy effect for perceived logical accuracy ($p < .001$), but not for confidence ($p > .05$). The significant source-condition effects on trust and overall evaluation also held (both $ps < .001$). 


\subsubsection{Cumulative Link Mixed Models}
As our dependent variables, perceived logical accuracy and confidence are ordinal instead of continuous, we fit cumulative link mixed models (CLMMs) in R with participant id, item id, and stimulus set id as random intercepts as an additional robustness analysis. For perceived logical accuracy, we used the model: clmm (Accuracy ~ Condition * Fallacy + (1 | Subject) + (1 | ItemID) + (1 | SetID), data = data). For confidence, we used the model: clmm (Confidence ~ Condition * Fallacy + (1 | Subject) + (1 | ItemID) + (1 | SetID), data = data). For both perceived logical accuracy and confidence, the substantive conclusions remained unchanged. However, as trust and overall evaluation scores were computed by aggregating multiple rating items, they were continuous rather than ordinal. Thus, we do not report the CLMM results for these variables. 

\textbf{Perceived Logical Accuracy.} For human participants, a significant Condition × Fallacy effect was obtained ($p < .001$) for perceived logical accuracy. Across all source conditions, participants rated comments containing logical fallacies as significantly less accurate compared to non-fallacious comments ($p < .001$).  Next, pairwise comparisons were run using the emmeans package in R with Tukey correction. Human and Human+AI conditions received significantly higher perceived logical accuracy ratings for fallacies compared to the other conditions (Control, AI, AI+Human), whereas non-fallacy comments received similar ratings across conditions. On the other hand, LLMs were stable across conditions, and no significant interaction effect was obtained. However, we obtained main effects for both fallacy presence and model. Logical fallacies were rated as less accurate than non-fallacies ($p < .01$). For non-fallacious comments, Gemini 2.5 Flash assigned the highest ratings, followed by GPT-5.2, and
then Claude Sonnet 4.5. For
fallacies, GPT-5.2 gave the highest ratings, followed by Gemini 2.5 Flash, and then Claude Sonnet 4.5.

\textbf{Confidence.} Human participants assigned similar confidence ratings for all labeled source conditions. However, they assigned lower confidence ratings to fallacies compared to non-fallacies in the Control condition ($p < .01$). LLMs were generally stable across source labels ($p > .05$), but differed in terms of models. Gemini 2.5 assigned the highest confidence scores, followed by GPT-5.2, and then Claude Sonnet 4.5.

\subsubsection{Robustness Across Open-Weight Models and Calibration-Based Confidence Measures}
We used self-reported Likert confidence for LLMs to enable a direct comparison with human evaluation, where confidence was also collected as self-reported Likert ratings. However, we additionally evaluated two open-weight models, Qwen-3.5-9B and Gemma-4-12B using calibrated probabilities and self-reported Likert confidence. Source labels had little effect on either model, similar to the proprietary LLM results. For Qwen, no pairwise differences existed after Holm correction for either self-reported or probability-derived confidence, regardless of the presence of fallacies. Gemma showed the same general pattern, with one exception: for fallacy-present items, its self-reported confidence was higher under the AI label than under Human+AI (4.92 vs. 4.75, $p_{adj}=.036$, $d=.44$). No other pairwise differences were observed for self-reported confidence or probability-derived confidence after correction. Thus, confidence was largely stable across source-label conditions, with only one isolated source-label effect for Gemma.

Self-reported and probability-derived confidence were positively associated, but the consistency of this relationship varied across models. For Qwen, correlations were moderate and significant across all source-label conditions for both fallacy-present ($\rho=.38$--$.52$, all Holm-adjusted $p\leq.011$) and fallacy-absent items ($\rho=.47$--$.60$, all Holm-adjusted $p<.001$), with similar associations when pooling across labels ($\rho=.46$ and $.53$, respectively). For Gemma, the relationship was weaker and less consistent: for fallacy-present items, significant positive correlations appeared for the no-label, Human+AI, and AI+Human conditions ($\rho=.35$--$.39$), but not for Human or AI labels, while the pooled association remained significant ($\rho=.31$, Holm-adjusted $p<.001$). For fallacy-absent items, only the no-label condition showed a significant correlation after correction ($\rho=.38$, Holm-adjusted $p=.042$), although the pooled association was small but significant ($\rho=.19$, Holm-adjusted $p=.020$). Overall, self-reported confidence tracked probability-derived confidence substantially more consistently for Qwen than for Gemma.

\end{document}